\documentclass[conference]{IEEEtran}
\usepackage[noadjust]{cite}
\usepackage{amsmath,amssymb,amsfonts}
\usepackage{algorithmic}

\usepackage{textcomp}

\usepackage{xspace}
\usepackage{graphicx}
\usepackage{color}

\usepackage{subcaption}
\definecolor{accessbl}{cmyk}{1, 0.4, 0.3, 0}
\DeclareCaptionFont{bl}{\color{accessbl}}
\usepackage{epsfig,endnotes,paralist,multirow}
\usepackage{wrapfig}
\usepackage{epsfig}
\usepackage[normalem]{ulem}
\usepackage{url}
\usepackage{soul}
\usepackage{multirow}
\usepackage{array}

\def\includeComments{include}
\def\includ{include}
\def\comm[#1]{\ifx\includeComments\includ  \bl \texttt{\textbf{\textit{Note: #1}}} \par \fi}
\def\inlinecomm[#1]{\ifx\includeComments\includ  \textit{Note: #1} \fi}

\newcommand{\red}[1]{\textcolor{black}{#1}}
\newcommand{\blx}[1]{\textcolor{black}{#1}}
\newcommand{\bleu}[1]{\textcolor{black}{#1}}
\newcommand{\blue}[1]{\textcolor{black}{#1}}
\newcommand{\tqbl}[1]{\textcolor{black}{#1}}

\newcommand{\para}[1]{\smallskip \noindent {\bf #1}}
\newcommand{\softpara}[1]{\smallskip \noindent \underline{#1}}
\newcommand{\vsp}{\vspace{0.01in}}

\newcommand{\cb}{\color{black}}
\newcommand{\cbl}{\color{black}}

\def\blackbox{\hfill {\vrule height6pt width6pt depth0pt}}

\newcounter{theorem}
\newtheorem{thm}{Theorem}

\newtheorem{corollary}{Corollary}

\newtheorem{defin}{Definition}
\newtheorem{ex}{Example}
\newtheorem{lem}{Lemma}
\newtheorem{ob}{Observation}

\newenvironment{theorem}{\vsp \begin{thm} \nopagebreak}{{\hfill$\blackbox$} \end{thm} \vsp}
\newenvironment{thm-prf}{\vsp \begin{thm} \nopagebreak}{\end{thm}}
\newenvironment{lemma}{\vsp \begin{lem} \nopagebreak}{{\hfill$\blackbox$} \end{lem} \vsp}
\newenvironment{lem-wo-prf-box}{\vsp \begin{lem} \nopagebreak}{\end{lem}}
\newenvironment{lem-prf}{\vsp \begin{lem} \nopagebreak}{\end{lem}}

\newenvironment{definition}[1]{\vsp\begin{defin}\begin{rm}(\textsc{#1})} {{\hfill$\Box$} \end{rm}\end{defin} \vsp}

\newenvironment{observation}{\noindent \begin{ob}}{{\hfill$\Box$}\end{ob}}
\newenvironment{cor-prf}{\vsp \begin{corollary} \nopagebreak}{\end{corollary}}

\newcommand{\R}{\ensuremath{\mathcal{R}}\xspace}
\newcommand{\T}{\ensuremath{\mathcal{T}}\xspace}

\newcommand{\clf}{\mbox{\tt Caleffi}\xspace}
\newcommand{\hh}{\ensuremath{h}}

\newcommand{\eps}{\mbox{EP}\xspace}
\newcommand{\epss}{\mbox{EPs}\xspace}
\newcommand{\es}{\mbox{ES}\xspace}
\newcommand{\ess}{\mbox{ESs}\xspace}
\newcommand{\os}{\mbox{\tt WaitLess}\xspace}
\newcommand{\wt}{\mbox{\tt Waiting}\xspace}
\newcommand{\DP}{\mbox{\tt DP}\xspace}
\newcommand{\dpa}{\mbox{\tt DP-Approx}\xspace}
\newcommand{\naive}{\mbox{\tt SP}\xspace}
\newcommand{\dpo}{\mbox{\tt DP-OPT}\xspace}
\newcommand{\iter}{\mbox{\tt ITER}\xspace}
\newcommand{\iterdpalt}{\mbox{\tt ITER-Bal}\xspace}
\newcommand{\iterdpa}{\mbox{\tt ITER-DPA}\xspace}
\newcommand{\iternaive}{\mbox{\tt ITER-SP}\xspace}
\newcommand{\qnr}{\mbox{\tt QNR}\xspace}
\newcommand{\spp}{\mbox{\tt QNR-SP}\xspace}
\newcommand{\LP}{\mbox{\tt LP}\xspace}
\newcommand{\dpalt}{\mbox{\tt Balanced-Tree}\xspace}
\newcommand{\qcast}{\mbox{\tt Q-Cast}\xspace}
\newcommand{\delftlp}{\mbox{\tt Delft-LP}\xspace}

\newcommand{\eat}[1]{}

\let\emph\textit

\newcommand{\pp}{\ensuremath{\bar{p}}}
\newcommand{\php}{\mbox{$p_{ob}$}\xspace}          
\newcommand{\bt}{\mbox{$t_{b}$}\xspace}         
\newcommand{\bp}{\mbox{$p_{b}$}\xspace}          
\newcommand{\gt}{\mbox{$t_g$}\xspace}      
\newcommand{\gp}{\mbox{$p_g$}\xspace}       
\newcommand{\ep}{\mbox{$p_e$}\xspace}       
\newcommand{\ct}{\mbox{$t_c$}\xspace}      

\newcommand{\fidl}{\ensuremath{\tau_l}\xspace}
\newcommand{\fidd}{\ensuremath{\tau_d}\xspace}

\newcommand{\ket}[1]{\ensuremath{|#1\rangle}}
\newcommand{\norm}[1]{\ensuremath{|#1|}}

\newcommand{\entT}[3]{
        \expandafter\ifx\expandafter\relax\detokenize{#1}\relax
            \expandafter\ifx\expandafter\relax\detokenize{#2}\relax
                \ensuremath{\mathrm{T}}\xspace
            \else
                \ensuremath{\mathrm{T_{#2,#3}}}\xspace
            \fi
        \else
            \ensuremath{\mathrm{T^#1_{#2,#3}}}\xspace
        \fi\xspace}
\newcommand{\tauT}[3]{
        \expandafter\ifx\expandafter\relax\detokenize{#1}\relax
            \expandafter\ifx\expandafter\relax\detokenize{#3}\relax
                \ensuremath{\mathrm{\tau}}\xspace
            \else
                \ensuremath{\mathrm{\tau_{#2,#3}}}\xspace
            \fi
        \else
            \ensuremath{\mathrm{\tau^#1_{#2,#3}}}\xspace
        \fi\xspace}

\def\BibTeX{{\rm B\kern-.05em{\sc i\kern-.025em b}\kern-.08em
    T\kern-.1667em\lower.7ex\hbox{E}\kern-.125emX}}

\begin{document}

\title{Efficient Quantum Network Communication using Optimized Entanglement-Swapping Trees}


\author{\IEEEauthorblockN{Mohammad Ghaderibaneh, Caitao Zhan, Himanshu Gupta, C. R. Ramakrishnan}
\IEEEauthorblockA{\textit{Department of Computer Science} \\
\textit{Stony Brook University, USA}\\
}
}

\maketitle

\begin{abstract}
Quantum network communication is 
challenging, as the No-cloning theorem in quantum regime
makes many classical techniques inapplicable; \tqbl{in particular, 
direct transmission of qubit states over long distances is infeasible
due to unrecoverable errors.}
For long-distance communication \tqbl{of unknown quantum states}, 
the only viable communication approach \tqbl{(assuming local operations
and classical communications)} is 
teleportation of quantum states, which requires a prior distribution of 
entangled pairs (\epss) of qubits.
Establishment of \epss across remote nodes can incur significant 
latency due to the low probability of success of the underlying 
physical processes. 
The focus of our work is to develop
efficient techniques that minimize \eps generation latency. Prior works
have focused on selecting entanglement \textit{paths}; in contrast, we\eat{propose
selection of efficient} \bleu{select}  \emph{entanglement swapping trees}---a more accurate 
representation of the entanglement generation structure. 
\eat{In this context,}\bleu{We} develop a dynamic programming algorithm 
to select an optimal
swapping-tree for a single pair of nodes, under the given capacity
and fidelity constraints. For the general setting,
we develop an 
efficient iterative algorithm \blue{to compute a set of swapping trees}.
We present simulation results which show that our solutions outperform 
the prior approaches by an order of magnitude and are viable for long-distance entanglement
generation.
\end{abstract}


\section{Introduction}
\label{sec:intro}

Fundamental advances in physical sciences and engineering have led to the realization of working quantum computers (QCs)~\cite{google-nature-19,ibm-quantum-roadmap}.  
However, there are significant limitations 
to the capacity of individual QC~\cite{Caleffi18}.  \tqbl{Quantum networks} (QNs) enable the construction of large, robust, and more capable quantum computing platforms by connecting smaller QCs. Quantum networks~\cite{qn}
also enable various important applications~\cite{qsn,qkd,atomic,secure,byzantine}.
However, quantum network communication is challenging --- e.g., physical transmission of quantum states across nodes can incur irreparable communication errors, as 
the No-cloning Theorem~\cite{Dieks-nocloning} proscribes making 
independent copies of arbitrary qubits. 
At the same time, certain aspects unique to the quantum regime, such as entangled states, enables 
novel mechanisms for communication.
In particular, teleportation~\cite{Bennett+:93} transfers quantum states with just classical
communication, but requires an a priori establishment of entangled pairs (\epss).
This paper presents 
techniques for efficient establishment of \epss in a network.

Establishment of \epss over long distances is challenging. 
Coordinated entanglement swapping (e.g. DLCZ protocol~\cite{dlcz}) using quantum repeaters 
can be used to establish long-distance entanglements from short-distance entanglements.
However, due to low probability of success of the underlying physical processes
(short-distance entanglements and swappings),  \eps generation can incur significant latency---
of the order of 10s to 100s of seconds between nodes 100s of kms away~\cite{gisin}.
Thus, we need to develop techniques that can facilitate fast generation of long-distance 
\epss. We employ two strategies to minimize generation latencies: (i) select optimal swapping 
{\bf trees} (not, just paths as in prior works~\cite{sigcomm20, delft-lp, guha, greedy2019distributed}) with a protocol that \bleu{retains unused \epss}; 
(ii) use multiple trees for each given node pair; this reduces effective latency by using
all available network resources.
In the above context, we address the following problems: 
(i) \spp Problem: Given a single $(s,d)$ pair, select a minimum-latency swapping tree under given constraints. 
(ii) \qnr Problem: Given a set of source-destination $(s,d)$ pairs, select a set of swapping trees for each pair with maximum aggregate \eps generation rate, under fidelity and resource constraints. 

To the best of our knowledge, no prior work has addressed the problem of selecting an efficient
swapping-tree for entanglement routing; they all consider selecting routing \textit{paths} \bleu{(\cite{caleffi} selects a path using a metric based on balanced trees; see \S\ref{sec:related}).}
Almost all prior works have considered the ``waitless''
model, wherein all underlying physical processes much succeed \textit{near-simultaneously} for 
an \eps to be generated; this model incurs minimal decoherence, but yields
very low \eps generation rates. 
In contrast, we consider the ``waiting'' protocol, wherein, at each swap operation, the earlier arriving 
\eps waits for a limited time for the other \eps to be generated. Such an approach with efficient swapping trees yields high entanglement rates; \bleu{the potential decoherence risk can be handled by discarding qubits that "age" beyond a certain threshold.}
\eat{at the cost of potential decoherence. However, qubit coherence times of the order of several 
minutes to a few hours~\cite{dec-2021,dec-15} have been demonstrated, 
which renders our approach feasible even for long distances.}

\para{Our Contributions.} 
We \eat{motivate formulation of} \bleu{formulate} the entanglement routing problem (\S\ref{sec:problem})
in QNs in terms of selecting optimal swapping \textit{trees} in the ``waiting'' protocol, under fidelity
constraints. In this context, we make the following contributions:
\begin{enumerate}
\item 
For the \spp problem, we design an optimal algorithm with
fidelity and resource constraints (\S\ref{sec:single-path}). 

\item
Though polynomial-time, the above optimal algorithm has high time complexity; \eat{we 
introduce some approximations, to improve the time-complexity of the algorithm
without degrading its empirical performance.}
we thus also design a near-linear time heuristic for the \spp problem based on an appropriate metric 
which essentially 
restricts the solutions to balanced swapping trees (\S\ref{sec:efficient}).

\item 
For the general \qnr problem,  we design an efficient iterative augmenting-tree algorithm (\S\ref{sec:iterative}), and show its effectiveness w.r.t.\ an optimal LP solution based on hypergraph-flows. 

\item 
We conduct extensive evaluations (\S\ref{sec:eval}) using NetSquid simulator, and show that our solutions outperform the prior approaches by an order of magnitude, while 
incurring 
little fidelity degradation. We also show that our schemes can generate high-fidelity
\epss over nodes 500-1000kms away.
\end{enumerate}

\eat{
\para{Overall Pitch and .}
Generate long-distance EPs “optimally” using a more “practical” model which makes use of memories (and transient storage of intermediate EPs).
Multiple paths, and Multiple Pairs. -----> Also, “optimally’.
Evaluation to:
Validate the realism of the high-level processes/protocols.
corroborate the expected superior “performance”.
}

\section{QC Background}
\label{sec:background}

\para{Qubit States.}
Quantum computation manipulates \emph{qubits}
analogous to how classical computation manipulates \emph{bits}.  
At any given time, a bit may be in one of two states, traditionally represented by $0$
and $1$.  A quantum state represented by a
\emph{qubit} is a \emph{superposition} of classical
states, and is usually written as $\alpha_0\ket{0} + \alpha_1\ket{1}$,
where $\alpha_0$ and $\alpha_1$ are \emph{amplitudes} represented by
complex numbers and such that $\norm{\alpha_0}^2 + \norm{\alpha_1}^2 = 1$.
Here, $\ket{0}$ and $\ket{1}$ are the standard (orthonormal) 
\emph{basis} states; concretely, they
may represent physical properties such as spin (down/up),
polarization, charge direction, etc.
When a qubit such as above is \textit{measured}, it collapses to a 
$\ket{0}$ state with a probability of $\norm{\alpha_0}^2$ and to
a $\ket{1}$ state with a probability of $\norm{\alpha_1}^2$.
In general, a state of an $n$ qubit
system can be represented as $\Sigma_{i=0}^{2^n-1} \alpha_i\ket{i}$
 where ``$i$'' in $\ket{i}$ is $i$'s bit representation. 
\eat{($\Sigma_{i=0}^{2^n-1}\norm{\alpha_i}^2=1$)}

\para{Entanglement.}
Entangled pure\footnote{\tqbl{In this work, we largely deal with only
pure qubit states. Entanglement of general mixed states is defined in
terms of separation of density matrices~\cite{qd-2009}.}}
states are multi-qubit states that cannot be
"factorized" into independent single-qubit states.
E.g., the 
$2$-qubit state $\frac{1}{\sqrt{2}}\ket{00}
+ \frac{1}{\sqrt{2}}\ket{11}$; this particular system is a
\emph{maximally-entangled} state. 
We refer to maximally-entangled pairs of qubits as \epss.
The surprising aspect of entangled states is that the combined
system continues to stay entangled, even when 
the individual qubits are physically separated by large distances.
This facilitates many applications, e.g., teleportation of qubit
states by local operations and classical information exchange, as
described next.

\begin{figure}
    \centering
    \includegraphics[width=0.49\textwidth]{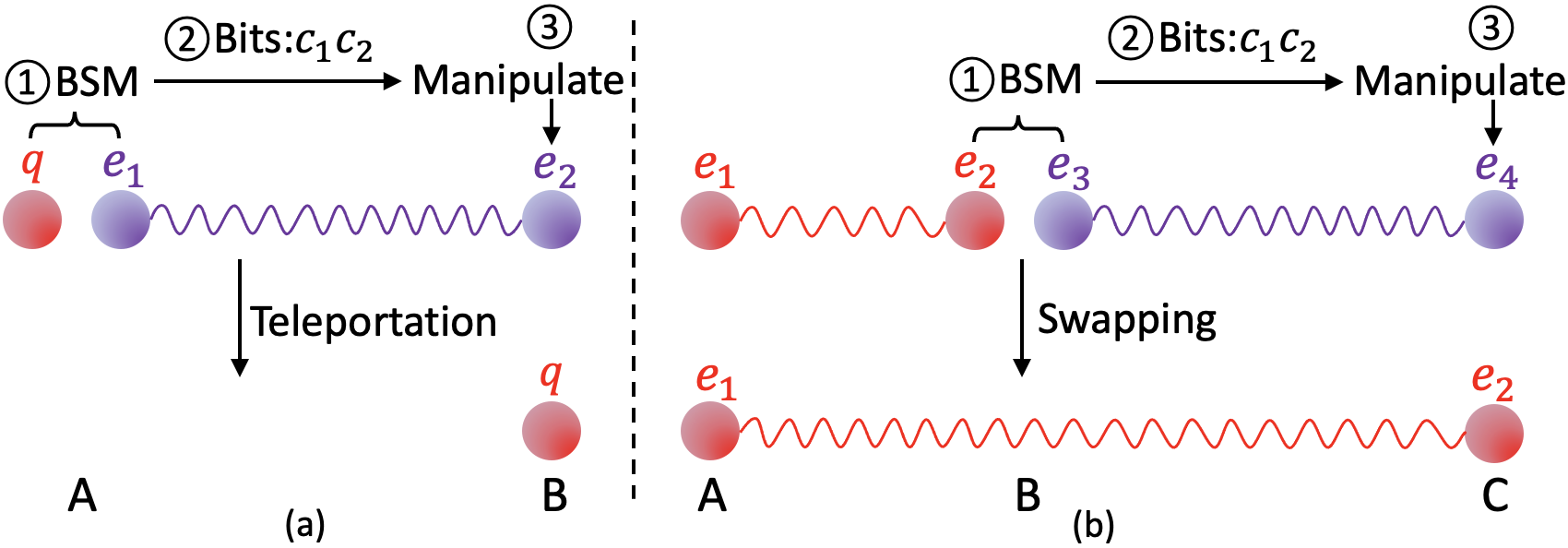}
    \vspace{0.1in}
    \caption{(a)  Teleportation of $\ket{q}$ from $A$ to $B$, while consuming an entangled pair $(e_1, e_2)$. (b) Entanglement swapping over the triplet of nodes $(A, B, C)$, which results in $A$'s qubit entangled with $C$'s qubit. This can be viewed as a teleportation of $e_2$ from node $B$ to $C$.}
  \label{fig:teleport_swap}
\end{figure}

\para{Teleportation.} 
Direct transmission of quantum data 
is subject to unrecoverable errors, 
as classical procedures such as amplified signals 
or re-transmission cannot be applied due to quantum no-cloning~\cite{wooterszurek-nocloning,Dieks-nocloning}.\footnote{\tqbl{Quantum error
correction mechanisms~\cite{muralidharan2016optimal, Devitt_2013} can be used to mitigate the transmission errors, 
but their implementation is very challenging and is not expected to be used
until later generations of quantum networks.}}
An alternative mechanism
for quantum communication is \emph{teleportation}, Fig.~\ref{fig:teleport_swap} (a), where a qubit $q$ from a node $A$
is recreated in another node $B$ (\emph{while ``destroying'' the original
qubit $q$}) using only classical communication.
However, this process requires that an \eps 
already established over the nodes $A$ and $B$. 
Teleportation can thus be used to reliably transfer quantum information.
At a high-level, 
the process of teleporting an arbitrary qubit,
say qubit $q$, from node $A$ to node $B$ can
be summarized as follows:
\begin{enumerate}
\item an \eps pair $(e_1, e_2)$ is generated over
$A$ and $B$, with  $e_1$ stored at $A$ and
$e_2$ stored at $B$;

\item 
at $A$, a \textit{Bell-state measurement} (BSM) operation 
over $e_1$ and $q$ is performed,
and the 2 classical bits measurement output $(c_1 c_2)$ is sent to $B$
through the classical communication channel; 
at this point, the qubits $q$ and $e_1$ at $A$ are destroyed.

\item manipulating the \eps-pair qubit $e_2$ 
at $B$ based on received $(c_1, c_2)$ 
changes its state to $q$'s initial state.
\end{enumerate}
Depending on the physical realization of qubits and the BSM operation, 
it may not always be possible to successfully generate the 2 classical bits,
as the BSM operation is stochastic.

\para{Entanglement Swapping (\es).} 
Entanglement swapping is an application of teleportation 
to generate \epss over remote nodes. See Fig. \ref{fig:teleport_swap} (b).
If $A$ and $B$ share an \eps  and $B$ teleports its qubit to $C$, then 
 $A$ and $C$ end up sharing an \eps. 
More elaborately, let us assume that $A$ and $B$
share an \eps, and $B$ and $C$ share a separate \eps. 
Now, $B$ performs a  BSM 
on its two qubits and communicates the result to $C$ (teleporting its 
 qubit that is entangled with $A$ to $C$). When $C$ finishes the protocol, it has a qubit
that is entangled with $A$'s qubit. Thus, an entanglement swapping (\es)
 operation can be looked up as being performed over a triplet of nodes $(A, B, C)$ 
with \eps available 
at the 
two pairs of adjacent nodes $(A, B)$ and $(B,C)$; it results in an \eps over the pair
of nodes $(A, C)$. 

\para{Fidelity: Decoherence and Operations-Driven.}
Fidelity is a measure of how close a realized state is to the ideal. 
Fidelity of qubit decreases with time, due to interaction with the environment,
as well as gate operations (e.g., in \es). 
Time-driven fidelity degradation is called \emph{decoherence}. 
To bound decoherence, we limit the aggregate time a qubit spends in a 
quantum memory before being consumed.
With regards to operation-driven fidelity degradation, 
Briegel et al.~\cite{BreigelEtAl1998} give an expression 
that relates the fidelity of an \eps generated by \es 
to the fidelities of the operands,  in terms of the 
noise introduced by swap operations and the number of link \epss used. The order of the swap operations (i.e., the structure of the 
swapping tree) does not affect the fidelity. 
Thus, the operation-driven fidelity degradation of the final \eps 
generated by a swapping-tree $T$
can be controlled by limiting the number of leaves of $T$, assuming that
the link \epss have uniform fidelity (as in~\cite{delft-lp}).

Entanglement Purification~\cite[e.g.]{BreigelEtAl1998} and Quantum Error 
Correction~\cite[e.g.]{QEC} have been widely used to combat fidelity 
degradation.  
Our work focuses on optimally scheduling \es operations with constraints
on fidelity degradation, without purification or error correction.

\para{Quantum Memories.}
Multiple quantum memories have been recently proposed to 
bring \tqbl{quantum networks} into realization. 
Types of quantum memories that support BSM measurements and gate unitary operations, and probably have a long decoherence time can be used in quantum communications.
Most of them are matter-based 
which have photonic interface to produce matter-matter entanglement 
over two neighboring nodes (see below).
At a high-level, there are three different quantum memory platforms: quantum dots, trapped atoms or ions, and colour centers in diamond. 
Each has its own physical characteristics and applications. 
While quantum dots have the ability to process quantum information very fast, 
they exhibit a very low decoherence time among 
others~\cite{press2010ultrafast, wang2019towards}. 
To overcome the low efficiency of single atom-photon coupling process, 
atomic ensemble schemes have been proposed~\cite{dlcz} where along with dynamic decoupling and cooling techniques, decoherence times of a few seconds have been achieved~\cite{sagi10, vetsch10, deutsch10}.
For trapped ion memories, decoherence times from several minutes to few hours have been demonstrated~\cite{ionqmemory05, ionqmemory21}.
To further 
increase the entanglement generation rate,~\cite{bhaskar2020experimental} proposes a way to use a single silicon–vacancy (SiV) 
colour center in diamond to perform asynchronous photonic BSM at 
the node located in the middle of two adjacent quantum nodes.

\subsection{Generating Entanglement Pairs (\epss)}
\label{sec:eps}

As described above, teleportation, which is the only viable means of transferring quantum states over long distances, requires an a priori distribution of \epss. 
Thus,   we need efficient mechanisms to  establish \epss across remote QN nodes; 
this is the goal of our 
work.
Below, we start with describing how \epss are generated between adjacent
(i.e., one-hop away) nodes, and 
then discuss how \epss across a pair of remote nodes can be established 
via \ess.

\para{Generating \eps over Adjacent Nodes.} 
The physical realization of qubits 
determines the technique used for sharing EPs between adjacent nodes. 
The \emph{heralded entanglement} process~\cite{sigcomm20, caleffi} to 
generate an atom-atom \eps between adjacent nodes $A$ and $B$
is as follows:
\begin{enumerate}
    \item 
Generate an entangled pair of atom and a telecom-wavelength photon at node $A$
and $B$.  
Qubits at each node are generally realized in an atomic form
for longer-term storage, while photonic qubits are used for transmission.

\item 
Once an atom-photon entanglement is locally generated
at each node (at the same time), 
the telecom-photons are then transmitted over an optical fiber to a photon-photon/optical BSM device $C$ located in the middle of 
$A$ and $B$ so that the photons arrive at $C$ 
at the same time.

\item 
The device $C$ performs a BSM over the photons, 
and transmits the classical result to $A$ or $B$ to complete \es.
\end{enumerate}
Other entanglement generation processes have been 
proposed~\cite{MuralidharanEtAl2016}; 
our techniques themselves are independent of how the link \eps are generated.



\begin{figure}
    \centering
    \includegraphics[width=0.48\textwidth]{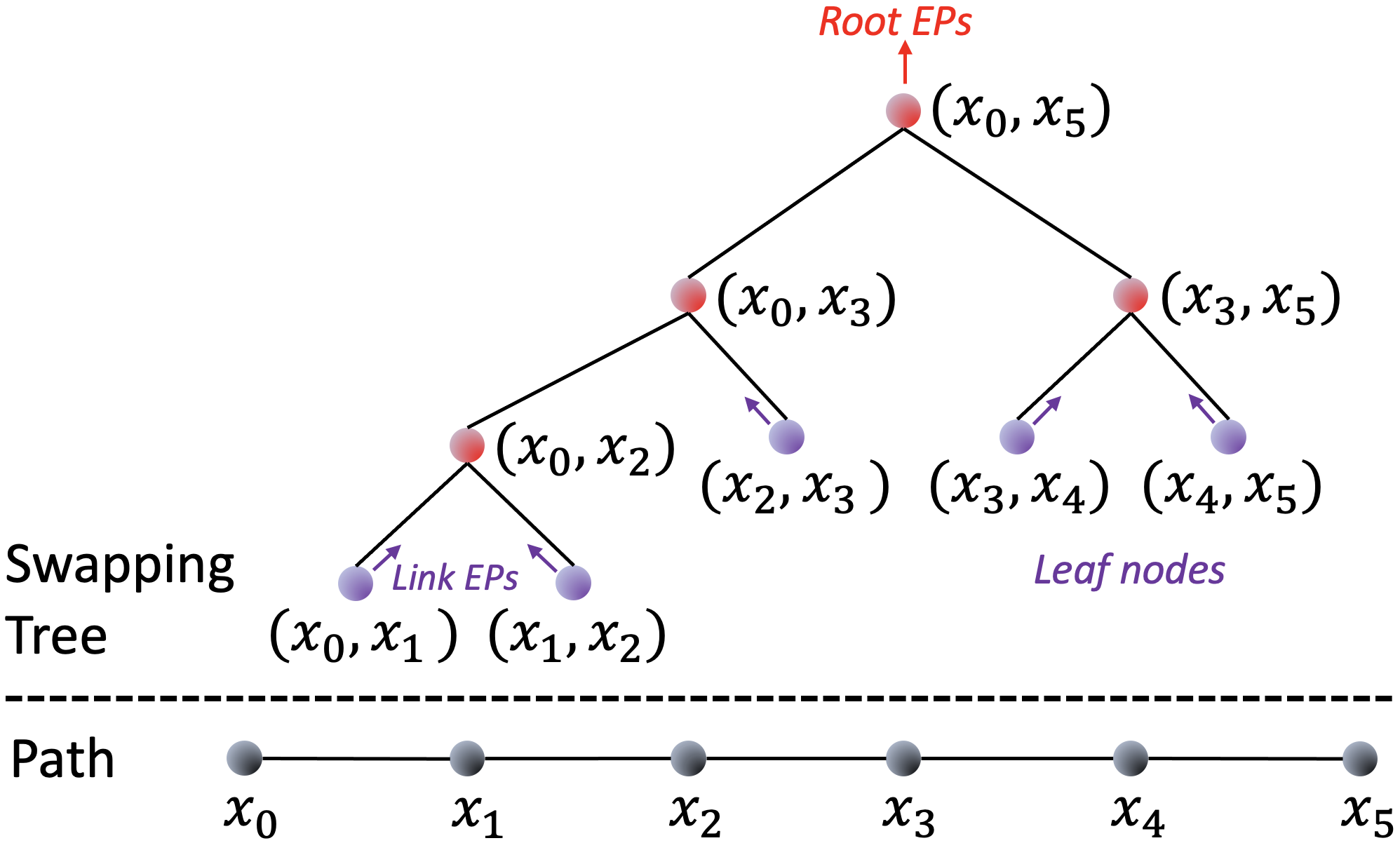}
    \vspace{0.1in}
    \caption{A swapping tree over a path. The leaves of the tree are the path-links, which generate link-\epss continuously.}
\label{fig:tree}
\end{figure}

\para{Generating \eps between Remote Nodes.} 
Now, \eps between non-adjacent nodes connected by a path in the network 
can be established by 
performing a sequence of \ess at 
intermediate nodes; this requires an a priori \eps over 
each of the adjacent pairs of nodes in the path.
For example, consider a path of nodes $x_0, x_1, x_2, x_3, x_4, x_5$, with an \eps 
between every pair of adjacent nodes $(x_i, x_{i+1})$. Thus, each node $x_i$
($1 \leq i \leq 4$) has two qubits, one of which is entangled with $x_{i-1}$ 
and the other with
$x_{i+1}$. Nodes $x_0$ and $x_5$ have only one qubit each.
To establish an \eps 
between $x_0$ and $x_5$, we can perform a sequence of entanglement
swappings (\ess) as shown in Fig.~\ref{fig:tree}. 
\eat{over the following triplet of nodes:
(i) \es over  $(x_1, x_2, x_3)$ to yield an \eps over $(x_1, x_3)$,
(ii) \es over  $(x_1, x_3, x_4)$ to yield an \eps over $(x_1, x_4)$,
(iii) \es over  $(x_1, x_4, x_5)$ to yield an \eps over $(x_1, x_5)$,
(iv) \es over  $(x_0, x_1, x_5)$ to yield an \eps over $(x_0, x_5)$.}
Similarly, the sequence of \es over the following triplets would also 
work: $(x_2, x_3, x_4)$, $(x_2, x_4, x_5)$, $(x_0, x_1, x_2)$, 
$(x_0, x_2, x_5)$. 
\eat{Note that some of the above \es can happen \ul{in parallel}, e.g.,
$(x_0, x_1, x_2)$ and $(x_2, x_3, x_4)$ 
can happen 
in parallel, as 
they are over separate sets of qubits.}

\begin{figure}
    \centering
    \includegraphics[width=0.49\textwidth]{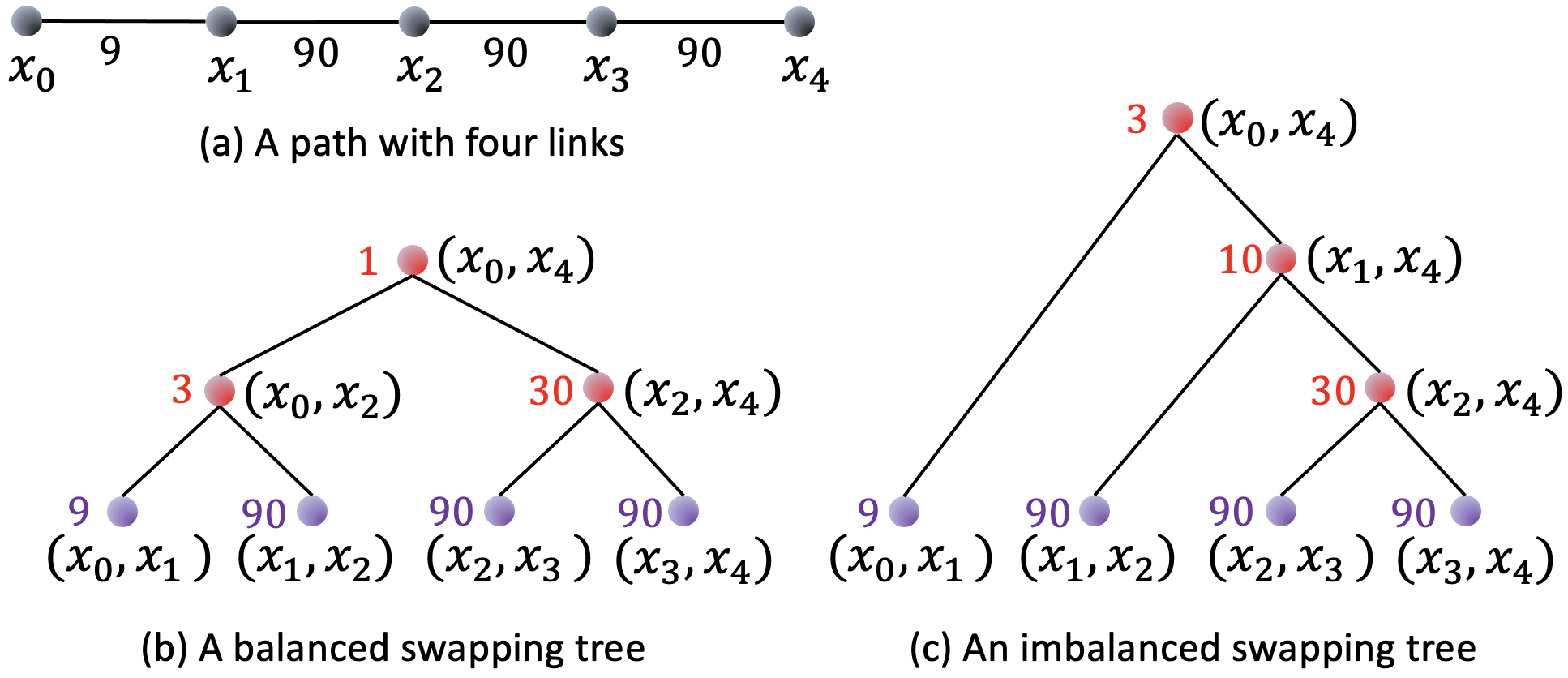}
    \vspace*{0.1in}
    \caption{Consider the path in (a). The imbalanced tree of (b) has a higher \eps generation rate than that of the balanced tree of (c). Here, the numbers represent the \eps generation rates over adjacent links or node-pairs.} 
    \vspace*{-0.2in}
    \label{fig:non-balance}
\end{figure}

\para{Swapping Trees.}
In general, given a path $P = s \leadsto d$ from $s$ to $d$, 
any complete binary tree (called a \textit{swapping tree}) over 
the ordered links in $P$ gives a way to generate an \eps over $(s, d)$.
Each vertex in the tree corresponds to a pair of network nodes in $P$, 
with each leaf representing a link in $P$. 
Every pair of siblings $(A, B)$ and $(B, C)$ perform an \es over 
$(A,B,C)$  to yield an \eps over $(A,C)$---their parent.
See Fig.~\ref{fig:tree}. 
\emph{Note that subtrees of a swapping tree execute in parallel.}
Different swapping trees over the same path $P$ can have different performance characteristics, as discussed later (see Fig.~\ref{fig:non-balance}).

\softpara{Expected Generation Latency/Rate of \epss.}
In general, our goal is to continuously generate \epss at some rate using 
a swapping tree, using continuously generated \epss at the leaves. 
The stochastic nature 
of 
\es operations
means that an \eps at the tree's root will be successfully 
generated only after many failed attempts and hence significant latency.
We refer to this latency as the \textit{generation latency} of the \eps at the 
root, and in short, just the generation latency of the tree. 
{\eps generation rate} 
is the inverse of its generation latency. Whenever we refer to generation 
latency/rate, we implicitly mean \textit{expected} generation latency/rate. 

\para{Two Generation Protocols: \os and \wt}
When a swapping tree is used to (continuously) generate \epss, there are two 
fundamentally different generation protocols~\cite{gisin,tittel-08}. 

\begin{itemize}
\item \textit{\os Protocol.}  
In this model, all the underlying processes, including link \eps generations and atomic
BSMs are synchronized. If all of them succeed then the end-to-end \eps is generated.
If \textit{any} of the underlying processes fail,
then all the generated \epss are discarded and the
whole process starts again from scratch (from generation of \eps at links).
In the \os protocol, all swapping trees over a given path $P$
incur the same generation latency---thus, here, the
goal is to select an optimal path $P$ (as in~\cite{sigcomm20,delft-lp}).  \eat{Since all underlying processes have to succeed simultaneously to generate an end-to-end EP, these protocols are inherently exponential in the number of nodes in $P$.} 

\item \textit{\wt Protocol.} In \wt protocol, a qubit of an \eps 
may wait (in a quantum memory) for its counterpart to become available so that an \es operation can be performed. 
Using such storage, we preclude discarding successfully generated \epss, and 
thus, reduce the overall latency in generation of a root-level \eps. 
E.g., let  $(A,B)$ and $(B,C)$ be two siblings in a swapping tree and \eps for $(A,B)$ is generated first. Then, \eps $(A,B)$ may wait for the \eps $(B,C)$ to be successfully generated. 
Once the \eps $(B,C)$ is generated, the \es operation
is done over the triplet $(A, B, C)$ to generate the \eps $(A,C)$. 
If the \eps $(A,B)$ waits beyond a certain threshold, then it may decohere.   
\end{itemize}

\softpara{Hardware Requirement Differences.}
\os protocols can generate \epss without quantum memories in a relay fashion if 
the EP generation among adjacent nodes can be tightly synchronized. In contrast,
\wt protocols benefit from memories with good coherence times (\S\ref{sec:eval}).

\para{Why \wt's Entanglement Generation Rate is Never Worse.} 
The focus of the \os protocol is to avoid qubit decoherence  
due to storage. But it results in very low generation rates
due to a very-low probability of \textit{all} the underlying 
processes succeeding at the same time. 
However, since qubit coherence times are 
typically higher than the link-generation latencies\footnote{Link generation latencies 
for 5 to 100km links range from about 3 to 350 milliseconds
for typical network parameters~\cite{caleffi}, while coherence times of few
seconds is very realistic (coherence times of several seconds~\cite{Longdell-2005,Fraval-2005} have been shown 
long ago, and more recently, 
even coherence times of several minutes~\cite{dec-13,dec-14} to a few hours~\cite{dec-15,dec-2021} 
have been demonstrated.},
an appropriately designed \wt protocol will always 
yield better generation rates \textit{without significantly compromising the fidelity} (see Theorem~\ref{thm:os-wt}). 
The key is to bound the waiting time to limit decoherence
as desired; e.g., in our protocol, we restrict to trees with high expected 
fidelities (\S\ref{sec:problem}), and discard qubits that
"age" beyond a threshold (\S\ref{sec:dec}).
\bleu{Both protocols use the same number of quantum memories (2 per node), though the \wt protocols will benefit from low-decoherence memories; other hardware requirements 
also remain the same.}

\eat{We note that both protocols require the \red{same number of quantum memories}---two 
per node---but \wt protocol benefit from low-decoherence memories; 
other hardware 
requirements are also same.} 

\begin{theorem}
Consider a quantum network, a path $P$, a swapping-tree $T$ over $P$, a \os protocol $X$, and a \wt protocol $Y$. Protocol $Y$ discards qubits that age (stay in  memory) beyond a certain threshold $\tau$ (presumably, equal to the coherence time). 
We claim that $Y$'s \eps generation rate will at least be that of $X$, 
irrespective of $\tau$ and $T$ (as long as it is over $P$), 
while ensuring that \epss generated by $Y$ are formed by non-decohered qubits and the operation-driven fidelity degradation of $Y$ \epss is same as $X$.
\label{thm:os-wt}
\end{theorem}

The above theorem suggests that \wt approach is always a better 
performing approach, 
irrespective of the decoherence time/limitations. See proof 
in Appendix~\ref{app:os-wt}.



\eat{
\textit{Paths vs.\ Trees.} When using \os protocol, all swapping trees over a given path $P$
incur the same generation latency. Thus, in the context of \os protocol, the
goal is to select an optimal path $P$ (as in~\cite{sigcomm20,delft-lp}). 
In this paper, we use the \wt protocol, and focus on the selection 
of efficient swapping trees 
with low generation latency
and bounded fidelity. The only other work, to the best
of our knowledge, that uses the \wt protocol, is~\cite{caleffi}.  
However, even that work focuses on selection of paths (in exponential time) and constructs balanced trees over the selected path.
}

\section{Model, Problem, and Related Works}
\label{sec:problem}

In this section, we discuss our network model, formulate the problem addressed, and discuss 
related work. 

\para{Network Model.} 
We denote a quantum network (QN) with a graph $G = (V, E)$,
with $V = \{v_1, v_2, \ldots, v_n\}$  and $E = \{(v_i, v_j)\}$
denoting the set of nodes and links respectively.
Pairs of nodes connected by a link are defined as \textit{adjacent} nodes. 
We follow the network model in~\cite{caleffi} closely.
Thus, each node has an atom-photon \eps generator with generation 
latency (\gt) and probability of success (\gp). Generation latency
is the time between successive attempts by the node to excite the 
atom to generate an atom-photon \eps; this implicitly includes the times for
photon transmission, optical-BSM latency, and classical acknowledgement.
\textit{For clarity of presentation} and without loss of generality,  
we assume homogeneous network nodes with same parameter values.
The generation rate is the inverse
of generation latency, as before.
A node's atom-photon generation capacity/rate 
is its aggregate capacity, and may be split across its incident links 
(i.e., in generation 
of \epss over its incident links/nodes).
Each node is also equipped with a certain number of atomic
memories to store the  qubits of the atom-atom \epss. 

\begin{figure}
    \centering
    \includegraphics[width=0.4\textwidth]{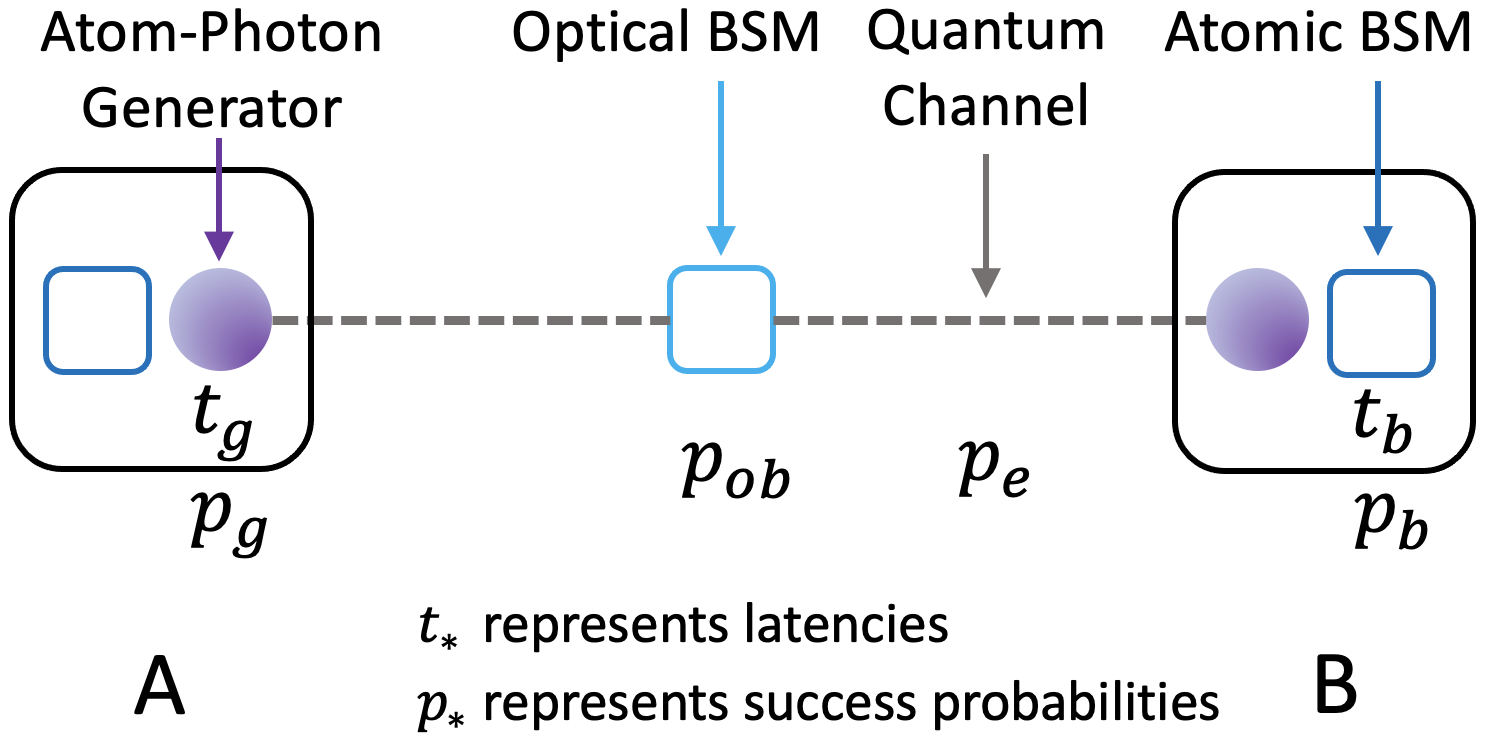}
  \vspace{0.1in}
  \caption{Key notations used.}
  \label{fig:notation_link}
\end{figure}
A network link is a quantum channel (e.g., using an 
optical fiber or a free-space link),
and, in our context, is used only for establishment 
of link \eps.
In particular, a link $e=(A,B)$ is used to transmit telecom-photons 
from $A$ and $B$
to the photon-photon BSM device in the middle of $e$.
Thus, each link is composed of two 
half-links with a probability of transmission success (\ep) that decreases exponentially with the link distance (see~\S\ref{sec:eval}).
The optical-BSM operation has 
a certain probability of success (\php).
To facilitate atom-atom \es operations, each network node is also equipped 
with an atomic-BSM device 
with an operation latency (\bt) and probability of success (\bp). Finally, 
there is an independent classical network with a transmission latency (\ct);
we assume classical transmission  
always succeeds.

\softpara{Single vs.\ Multiple Links Between Nodes.}   For our techniques multiple links between a pair of  adjacent nodes can be replaced by a single link of aggregated
rate/capacity.  Hence we
assume only a single link between every pair of nodes. However, 
distinct multiple links between nodes have been used creatively
in~\cite{sigcomm20} (which refers to them as multiple channels); thus, we will 
discuss multiple links further in~\S\ref{sec:eval} when we evaluate various techniques. \red{We note that the all-photonic protocol in~\cite{all-photo-15} is essentially a more sophisticated version of the multi-link \os protocol in~\cite{sigcomm20} to further
minimize memory requirements, but it uses multipartite cluster states which are
challenging to create. In either case, in terms of selection of paths/trees,
the path-selection techniques from~\cite{sigcomm20} should also apply to the all-photonic protocol with certain modifications to account for 
how the cluster states are generated.}

\para{\eps Generation Latency of a Swapping Tree.} 
Given a swapping tree and \eps generation rates at the leaves (network links), we 
wish to estimate the generation latency of the \epss over the remote pair corresponding
to the tree's root with the \wt protocol. 
Below, we develop a recursive equation.
Consider a node $(A,C)$ in the tree, with $(A,B)$ and $(B,C)$ as its two children. 
Let $T_{AB}, T_{BC}$, and $T_{AC}$ be the corresponding (expected) generation latencies 
of the \epss over the three pairs of nodes. Below, we derive an expression for $T_{AC}$
in terms of  $T_{AB}$ and $T_{BC}$; this expression will be sufficient to determine the
expected latency of the overall swapping tree by applying the expression iteratively.
We start with an observation.
\begin{observation}
If two \eps arrival processes $X_1$ and $X_2$ are exponentially distributed 
with a mean inter-arrival
latency of $\lambda$ each, then the expected inter-arrival latency of 
$\max(X,Y)$ is $(3/2)\lambda$.
\label{ob:expdist}
\end{observation}
From above, if assume $T_{AB}$ and  $T_{BC}$ to be exponentially 
distributed  
with the same expected generation latency of $T$, 
then the expected latency of both \epss arriving is $(3/2)T$. 
Thus, we have: 
\begin{equation}
T_{AC} = (\frac{3}{2} T + \bt + \ct)/\bp, \label{eqn:tree-rate}
\end{equation}
\softpara{Remarks.}
We make the following remarks regarding the above expression.
First, when $T_{AB} \neq T_{BC}$, we are able to only derive an upper-bound on 
$T_{AC}$ which is given by the above equation but with $T$ replaced by
$\max(T_{AB}, T_{BC})$.\footnote{\tqbl{The 3-over-2 formula as an upper bound has also
been corroborated in a recent work~\cite{analytical22} which derives analytical bounds 
on \eps latency times in more general contexts.}}
However, in our methods, the above assumption of $T_{AB} = 
T_{BC}$ will hold as we would only be considering ``throttled'' trees to save on 
underlying network resources (see~\S\ref{sec:single-path}).
\tqbl{Second, our motivation for the exponential distribution assumption stems from the
fact that the \eps generation latency at the {\em link level} is certainly exponentially
distributed if we assume the underlying probabilistic events to have a Poisson distribution.}
Third, note that
the resulting distribution is not exponential. 
Despite this, we apply the above equation recursively to compute the tree's generation
latency. \blue{However,} in our evaluations, we
observe the validity of this approximation since our analysis matches closely with the simulation results. 
Finally, Eqn.~\eqref{eqn:tree-rate}
is conservative in the sense that 
each round of an \eps generation of any subtree's root starts from 
\tqbl{scratch (i.e., with no link \epss from prior round) }
and ends with either a 
\eps generation at the \textit{whole swapping tree}'s root
or an atomic-BSM failure at the subtree's root.
We do not ``pipeline'' any operations across rounds within a subtree, which may lower
latency; this is beyond this work's scope. 

\eat{
Lastly, in the \os protocol, the expected generation 
latency is much higher due to the low-probability of \emph{all} 
of the underlying processes succeeding at the same time/time-slot. 
In addition, the \os protocol
requires that the underlying links be synchronized (which is not required
in the above \wt protocol). The expected generation latency in the \os protocol
is simply 
\begin{equation}
\frac{\gt}{(\gp^2\ep^2\php)^l\bp^l},  
\end{equation}
if we assume the link-\eps generation latency is uniformly \gt. Above, the path
length is $l$. When we have multiple links between adjacent nodes, then the expression
for generation latency is more complex as derived in~\cite{sigcomm20}.}

\subsection{Problem Formulation}
\label{sec:formulation}

We now formulate the central problem of selecting
\textit{multiple} swapping trees for each given source-destination pair. Selection of multiple
routes is a well-established strategy~\cite{sigcomm20, delft-lp, guha} to 
maximize entanglement rates.

\para{Quantum Network Routing (\qnr) Problem.}
Given a quantum network and a set of source-destination pairs $\{(s_i, d_i)\}$, 
the \qnr problem is to determine a set $\T_i$ of swapping trees for each pair $(s_i, d_i)$ 
such that the sum of the \eps rates 
of all the trees in $\bigcup_i \T_i$ is maximized under the following 
constraints: 
\begin{enumerate}
    \item \textit{Node Constraints.} For each node, the aggregate resources used by $\bigcup_i \T_i$ is less than the available resources; we formulate this formally below.
    \item \textit{Fidelity Constraints.} Each swapping tree in $\bigcup_i \T_i$ satisfies the following: (a) Number of leaves is less than a given threshold $\fidl$; this is to limit fidelity degradation due to gate operations. (b) Total memory storage time of any qubit is 
    less\footnote{\tqbl{We note that, in our context, the storage time as well as the memory coherence 
    time are statistical quantities due to the underlying statistical mechanisms. However, for the purposes of {\em selecting} a swapping tree, we use a fixed decoherence threshold \fidd\ value equal to the mean of the distribution of the coherence time (recent work~\cite{boxi-2020} computes optimal cut-offs/thresholds, and their techniques can be used to pick \fidd). When simulating a selected tree for generation of EPs, we can implement coherence time as a statistical measure.}}
    than a given \textit{decoherence threshold} $\fidd$.
\end{enumerate}
Informally, the swapping-trees may also satisfy some fairness constraint across the given 
source-destination pairs. A special case of the above \qnr problem is to select a single tree
for a source-destination pair; we address this in the next section. 

\softpara{Formulating Node Constraints.} 
Consider a swapping tree $\T \in \bigcup_i \T_i$  over a path $P$. For each link
$e \in P$, let $R(e, \T)$ be the \eps rate  being used by \T over the link $e$ in $P$. 
Let us define $R_e = \sum_\T  R(e,\T)$, and let $E(i)$ be the set of edges incident on $i$.
Then, the node capacity constraint is formulated as follows.
\begin{eqnarray}
1/\gt &\geq& \sum_{e \in E(i)} R_e/(\gp^2\ep^2\php) \ \ \ \  \forall i \in V. \label{eqn:qnr-1}
\end{eqnarray}
\tqbl{The above comes from the fact that to generate a single link \eps over $e$, each end-node of $e$ needs
to generate $1/(\gp^2 \ep^2 \php)$ photons successfully, since each photon (from each end-node) 
has a generation success of \gp and a transmission success rate of \ep, and the optical-BSM's success
probability over the two successfully arriving photons is \php. Note that $1/\gt$ is a 
node's total generation capacity.}
Also, the memory constraint is that for any node $i$, the memory available in $i$ should be more than $2x + y$ where $x$ is the number of swapping trees that use $i$ as an intermediate node 
and $y$ is the number of trees that use $i$ as an end node.

\subsection{Related Works}
\label{sec:related}

There have been a few works in the recent years that have addressed generating long-distance \epss
efficiently. All of these works have focused on selecting an efficient routing path for the swapping
process \bleu{(\cite{caleffi} also selects a path, but using a metric based
on balanced trees). }
In addition, all except~\cite{caleffi} have looked at the \os protocol of generating the
\epss. Recall that in the \os model, selection of paths suffice, while in the \wt model, one needs
to consider selection of efficient swapping trees with high fidelity.
Selection of optimal swapping trees is a fundamentally more challenging problem 
than selection of paths---and has not 
been addressed before, to the best of our knowledge. \eat{Efficient use of intermediate \epss rather than
discarding them (as in the \os protocol) has been mentioned as an open problem in~\cite{sigcomm20};
our paper addresses this challenge by using the \wt protocol.} We start with discussing how the \os model
works.

\para{\os Approaches.}
The most recent works to address the above problem are~\cite{sigcomm20} and~\cite{delft-lp}, 
both of which consider the \os model. 
In particular, Shi and Qian~\cite{sigcomm20} design a Dijkstra-like algorithm to construct an optimal path
between a pair of nodes, when there are multiple links (channels) between adjacent nodes. Then,
they use the algorithm iteratively to select multiple paths over multiple pairs of nodes.
Chakraborty et al.~\cite{delft-lp} design a multi-commodity-flow
like LP formulation to select routing paths for a set of source destination pairs. They map
the operation-based fidelity constraint to the path length (as in~\cite{BreigelEtAl1998}), and use
node copies to model the constraint in the LP. However, they explicitly assume that the link
\eps generation is deterministic---i.e., always succeeds. 
Among earlier relevant works,~\cite{guha} proposes a greedy solution
for grid networks, and~\cite{greedy2019distributed} proposes 
virtual-path based routing in ring/grid networks.
\eat{and~\cite{gradient} using a gradient approach to select
efficient routing paths.}

\para{\wt Approach.}
Due to photon loss, establishing long-distance entanglement between remote
nodes at $L$ distance by \textit{direct} 
transmission yields \eps rates that decay exponentially with $L$. 
DLCZ protocol~\cite{dlcz,gisin} broke this exponential barrier using
$2^k$ equidistant intermediate nodes to perform entanglement-swapping operations, 
implicitly over a balanced binary tree, with a \wt protocol; this makes the 
\eps generation rate decay only polynomially in $L$. \eat{This is fundamental to the feasibility of establishing long-distance entanglements.}
More recently, Caleffi~\cite{caleffi} formulated the entanglement generation rate on a given path between two nodes, under the more realistic condition where the intermediate nodes in the path may not all be equidistant, but still considered only balanced trees. Their path-based metric
was then used to select the optimal path by enumerating over the 
exponentially many paths in the network.

\softpara{Our Approach (vs.~\cite{caleffi}).}
Though~\cite{caleffi} considers only balanced trees, its 
brute-force algorithm is literally impossible 
to run for 
networks more than a few tens of nodes (\S\ref{sec:eval}).
In our work, we observe that a path has many swapping trees, 
and, in general, imbalanced trees may even
be better; see Figure~\ref{fig:non-balance}. 
Thus, we design a polynomial-time dynamic programming (DP) algorithm that delivers 
an \textit{optimal} high-fidelity swapping-tree;
our DP approach effectively considers all possible 
swapping trees, not just balanced ones (note that, even over a single path, 
there are exponentially many trees). Incorporation of fidelity (including decoherence) in our DP approach requires non-trivial observation and analysis (\S\ref{sec:dec}).
Our \dpalt Heuristic (\S\ref{sec:efficient}) is closer to~\cite{caleffi}'s work, 
in that both 
consider only balanced trees; however, we use a heuristic metric that facilitates a polynomial-time Dijkstra-like heuristic to select the optimal path, while their recursive metric~\footnote{We note that their formula (Eqn.~10 in~\cite{caleffi}) is incorrect as it either ignores the 3/2 factor or assumes the \eps generations to be synchronized {\bf across all} links. In addition, their expression for "qubit age" ignores the "waiting for \es" time completely. \label{ft:wrong}} 
(albeit more accurate than ours) is not amenable to an efficient (polynomial-time) search algorithm. 

\para{Other Works.}
In~\cite{Jiang17291}, Jiang et al.\ address a related problem; given a 
path with uniform link-lengths, they give an algorithm for selecting an 
optimal sequence of swapping and purification operations 
to produce an \eps with fidelity constraints.  
In other recent works, Dahlberg et al~\cite{sigcomm19} design physical and link layer protocols
of a quantum network stack, and~\cite{conext20} proposes a data plane protocol to generate \epss
within decoherence thresholds along a \emph{given} routing path. 
More recently, Bugalho et al.~\cite{bugalho2021distributing} propose an algorithm to efficiently distribute multipartite entanglement across over than two nodes.

\section{Optimal Algorithm for Single Tree}
\label{sec:single-path}

In this section, we consider a special case of the \qnr problem, viz., the case wherein there is a single source-destination $(s,d)$ pair and the goal is to select a \textit{single} swapping tree for the $(s,d)$ pair. 
For this special case, we design an optimal algorithm based on dynamic programming. 
This optimal algorithm can be used iteratively to develop an efficient heuristic for the general \qnr problem, as in~\S\ref{sec:iterative}.  

\para{\qnr \red{Single Path} (\spp) Problem.} Given a quantum network and a source-destination pair $(s,d)$, 
the \spp problem is to determine a single swapping tree that maximizes the expected
generation rate (i.e., minimizes the expected generation latency) of \epss over
$(s,d)$, under the capacity and fidelity constraints.

For homogeneous nodes and link parameters, it is easy to see that the best swapping-tree is the balanced or almost-balanced tree over the shortest path.
We note that \spp is not a special case of \qnr in the formal sense; e.g., 
the LP algorithm (\S\ref{sec:multiple-path}) for \qnr cannot be used for the \spp problem, due to the single
tree requirement (LP may produce multiple trees).
As described in \S\ref{sec:related}, the \spp problem has been addressed before in~\cite{sigcomm20, caleffi} under different models. 

\eat{E.g.,~\cite{sigcomm20} consider the problem under the \os model with multiple links
between adjacent nodes. Also,~\cite{caleffi} gives an exponential-time optimal algorithm
for the \spp problem restricted to only fully-balanced swapping trees; incidentally,
removing the restriction of fully-balanced trees allows us to 
design an optimal algorithm. }

\subsection{Dynamic Programming (\DP) Formulation}

First, we note that a Dijkstra-like shortest path approach which builds a shortest-path tree greedily doesn't work for the \spp problem---mainly, because the task is to find an optimal \textit{tree} rather than an optimal path. As noted before, a routing path can have exponentially many swapping trees over it,  with different generation latency. 
The recursive expression for computing the generation latency given in \S\ref{sec:problem} suggests that a dynamic programming (DP) approach, similar to the Bellman-Ford or Floyd-Warshall's classical algorithms for shortest paths,
may be applicable for the \spp problem. However, we need to ``combine'' trees
rather than paths in the recursive step of a DP approach. Consequently, we were unable to design a DP approach based on the Floyd-Warshall's approach, but, are able to extend the Bellman-Ford approach for the \spp problem after addressing a few challenges discussed below.

\para{DP Formulation.} 
We start with designing a DP algorithm without worrying about the decoherence 
constraint; we incorporate the decoherence constraint in the next subsection.
Given a network, let $T[i,j,\hh]$  be the optimal expected latency 
of generating \eps pairs over $(i,j)$ using a swapping tree of height 
at most \hh. 
Note that $T[i,j,0]$ for adjacent nodes $(i,j)$ can be given by 
$\frac{\gt}{\gp^2\ep^2\php}$.
Now, based on Eqn.~\eqref{eqn:tree-rate},
we start with the following 
equation for computing $T[i,j,\hh]$ in terms of smaller-height swapping trees. 

\begin{align}
    T[i,j,\hh] &= \min\big(T[i,j,\hh-1], (\frac{3}{2}  B + \ct + \bt)/\bp\big)  \label{eqn:simple-dp} \\
    {\rm where:} & \nonumber \\
    B &= \min_{k \in V}\ \ \max \big(T[i, k, \hh-1], T[k, j, \hh-1]\big) \nonumber
\end{align}

However, there are three issues that need to be addressed 
before the above formulation can be turned into a viable 
algorithm. We address these in the below three paragraphs. 

\softpara{(1) The 3/2 Factor; Throttled Trees.}
As mentioned in \S\ref{sec:problem}, the 3/2 factor is an accurate estimate if the corresponding $T$'s are equal. However, in the above equation, $T[i, k, \hh-1]$ and $T[k, j, \hh-1]$ may not be equal. 
In our overall methodology, to conserve node and link resources, we post-process or "throttle" the swapping-tree obtained from the DP algorithm by increasing the generation latencies of some of the {\bf non-root} nodes such that (i) the latencies of siblings are equalized, and (ii) the parents latency is related to the children's latency by Eqn.~\eqref{eqn:tree-rate}.
We refer to this post-processing as \textit{throttling}, and a tree that satisfies the above conditions as a \textit{throttled} tree. Note that throttling does not alter the generation latency of the root  and thus the overall tree; we prove the optimality of the overall algorithm formally in Theorem~\ref{thm:dp}. Below, we motivate throttling, and describe how it is achieved.

\textit{Justification.} In a given swapping tree, 
consider a pair of siblings $x$ and $y$ that have unequal generation 
latencies/rates. Let $x$ be the one with a lower latency (higher rate). Then, 
$x$ will likely have to discard many \epss while waiting for an \eps from $y$. 
To minimize this discarding of \epss from $x$ and to conserve underlying network 
resources so that they can be used in other swapping trees (in a general \qnr solution), 
we ``throttle'', increase (decrease) the generation latency (rate) of, the sibling $x$ 
to match that of $y$. 

\textit{Throttling Process.}
Consider a pair of siblings $x$ and $y$ in the tree; let their parent be $z$.
Let $T_x, T_y,$ and $T_z$ be their current generation latencies, such that 
$T_z = (\frac{3}{2} \max (T_x, T_y) + \ct + \bt)/\bp$. 
There are two potential steps: (i) If the parent's latency is to be kept 
unchanged, but $T_x < T_y$ then $T_x$ is increased to $T_y$ which, thus, makes the 
above equation valid. (ii) If the parent's latency $T_z$ is increased to $T$ (by
the above first step, with $z$ as a sibling), then we increase the latencies of 
both $x$ and $y$ to $2/3 (T \bp - \ct -\bt)$.
It is easy to see that applying the two steps iteratively from the root to the leaves, yields a throttled tree, as defined above. 

\softpara{(2) Capacity Violation at Node $k$.}
Note that the middle/common node $k$ in 
Eqn.~\eqref{eqn:simple-dp}
may violate
(node) capacity constraints in the merged tree corresponding to 
$T[i, k, \hh]$, as it may use
its full capacity in the trees corresponding to $T[i, k, \hh-1]$ and $T[k, j, \hh-1]$. 
We address the above by adding two additional parameters to the sub-problems \tqbl{function $T$},
corresponding to ``usage percentage'' of the end nodes.
In particular, we define $T[i,j,\hh,u_i,u_j]$ as the optimal latency of a swapping tree
of height at most \hh, under the constraint that the end nodes $i$ and $j$ use at most
$u_i$ and $u_j$ percentage of the respective node generation capacities; here, 
$u_i$ and $u_j$ can be positive integers between 1 and 100. 
The base case $T[i,j,0,u_i,u_j]$ for adjacent nodes $(i,j)$ is given by $\frac{\gt\min(u_i, u_j)}{\gp^2\ep^2\php}$.
Eqn.~\eqref{eqn:simple-dp}
is modified as follows to accommodate the
additional usage parameters.

\begin{align} \label{eqn:dp-usage} 
    T[i, j, \hh, u_i, u_j] = \min\big( 
    \!\begin{aligned}[t]
        &T[i, j, \hh-1, u_i, u_j],\hspace*{0.55cm}& \\
        &(\frac{3}{2} B + \ct + \bt)/\bp\big) 
    \end{aligned}\\
where: \hspace*{6.1cm} &\nonumber \\
     B = \min_{k,\ u + u' = 100} \max\big( \nonumber 
    \!\begin{aligned}[t]
        &T[i, k, \hh-1, u_i, u],\hspace*{0.75cm}& \\
        &T[k, j, \hh-1, u', u_j]\big)
    \end{aligned}
\end{align}


\softpara{(3) Ensuring Disjoint Subtrees.}
Note that Eqn.~\eqref{eqn:simple-dp}
implicitly assumes that the swapping trees corresponding 
to the latency values $T[i,k,\hh-1]$ and $T[k,j,\hh-1]$ are over disjoint paths, i.e., 
there is no node $v$ such that both the paths contain $v$. If there is a common node $v$, 
then the combined tree
corresponding to $[i,j,\hh]$ may violate the node capacity constraints at $v$. 
This issue also arises in the classical Bellman Ford's or Floyd-Warshall's algorithms for shortest
weighted paths, but is harmless with the assumption of positive-weighted cycles.
We resolve the issue similarly here via the below lemma (see Appendix~\ref{app:subtrees} for the proof).

\begin{lemma}
Consider two swapping trees $\T_{ik}$ and $\T_{kj}$ each of height at most $\hh -1$ 
over paths  $P_1: i \leadsto v \leadsto k$ and 
$P_2: k \leadsto v \leadsto j$, each of which contains a common node $v\not=k$.
Then there exists two swapping trees $\T_{iv}$ and $\T_{vj}$ each of height at most $\hh -1$
over paths $P'_1: i \leadsto v$ and $P'_2: v \leadsto j$ such that: (i) $P'_1$ is a subset of $P_1$, 
and $P'_2$ is a subset of $P_2$, and (ii) generation latency of $\T_{iv}$ is no greater than that of $\T_{ik}$, 
and generation latency of $\T_{vj}$ is no greater than that of $\T_{kj}$.
\label{lem:subtrees}
\end{lemma}

Lemma 1 implies that if the swapping trees $\T_{ik}$ and $\T_{kj}$ corresponding 
to the latency values $T[i,k,\hh-1]$ and $T[k,j,\hh-1]$ have a common node, then there exist 
swapping trees of equal or better latency without any common nodes and these trees can be used to build a lower-latency tree over $(i,j)$.

\para{Overall DP Algorithm and Optimality.} 
Our DP-based algorithm for the \spp problem for a given $(s,d)$ pair is as follows.
We use a DP formulation based on Eqn.~\eqref{eqn:dp-usage}
and the corresponding base case values 
to compute optimal generation latency $T[s, d, \hh, 100, 100]$ and the corresponding swapping tree
$\T$. Then, we throttle the tree $\T$ as described in paragraph (1) above. The below theorem \tqbl{(see
Appendix~\ref{app:dp} for proof)} states
that the throttled tree thus obtained has the optimal (minimum) expected generation latency among all
throttled trees.

\begin{theorem}
    The above described DP-based algorithm returns a throttled swapping tree over $(s,d)$ 
    with minimum expected generation
    latency (maximum expected generation rate) among all 
    throttled trees over the given $(s,d)$ pair.
    \label{thm:dp}
\end{theorem}

\subsection{Incorporating Fidelity Constraints}
\label{sec:dec}

Till now, we have ignored the fidelity constraints. We incorporate
them in this section, by extending our DP formulation from the previous section.
Limiting the decoherence, i.e., the qubit storage time, is challenging and 
is addressed first below. Limiting the number of leaves of a swapping tree 
is relatively easier, and is discussed next. We start with a definition.

\begin{definition}{Qubit/Tree Age.}
Given a swapping tree, the total time spent by a qubit in a swapping
tree is the time spent from its ``birth'' via an atom-photon \eps generation at a node till its consumption
in a swapping operation or in generation of the tree's root \eps. 
We refer to this as a qubit's \textit{age}. The maximum age over all qubits 
in a swapping
tree is called the tree's (expected) \textit{age}. 
\end{definition}

\begin{figure}
    \centering
    \includegraphics[width=0.44\textwidth]{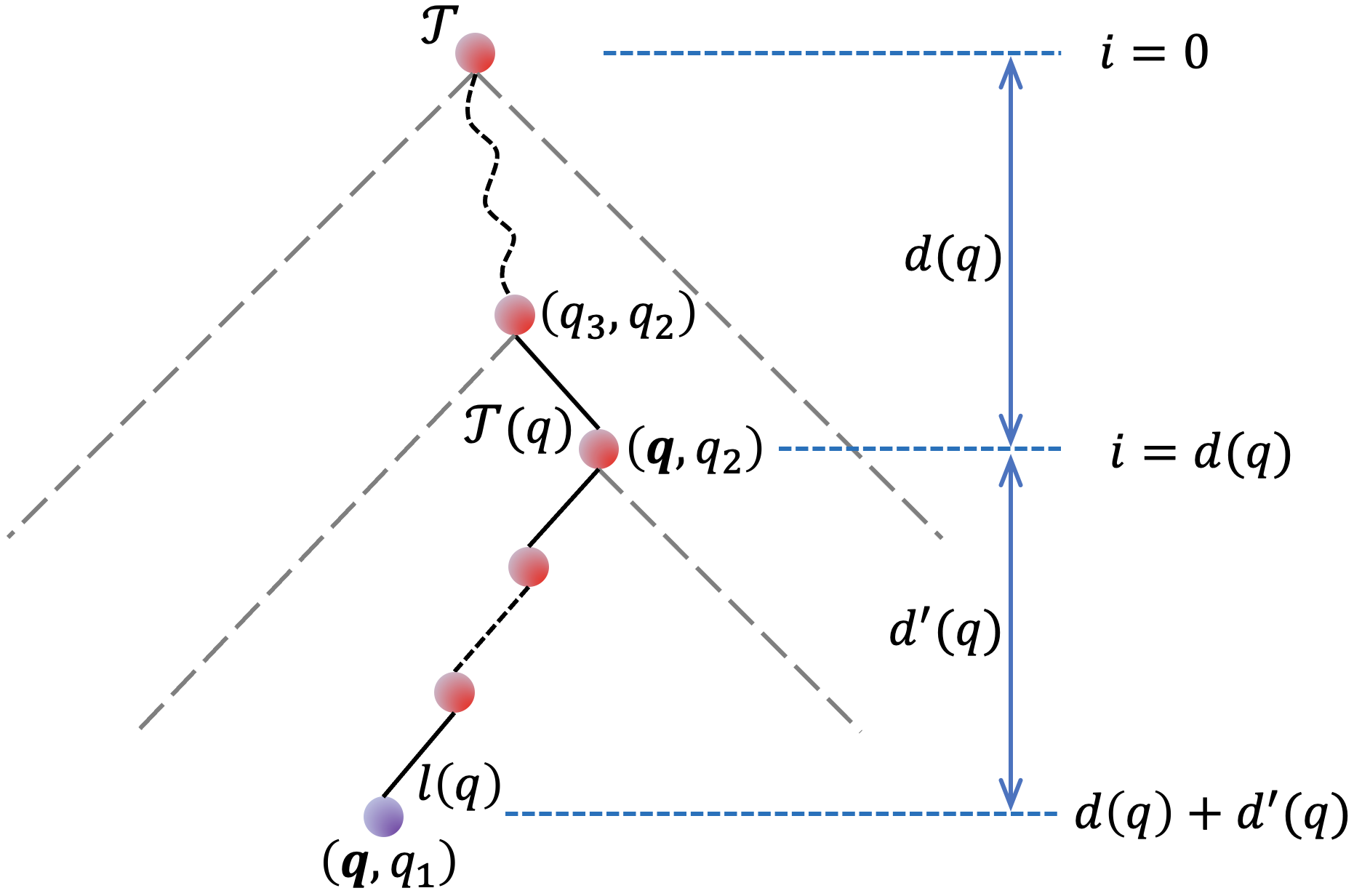}
    \vspace{0.1in}
  \caption{Qubit parameters in a swapping tree used to compute the \emph{age} of a qubit $q$ at a leaf node $l(q)$. Here, $l(q)$ is the left-most leaf of the subtree $\T(q)$.}
  \label{fig:age}
\end{figure}

\para{Estimating Qubit Age in a Swapping Tree.}
Consider a throttled swapping tree $\T$, with a generation latency of $T$.
Consider two siblings $(A,B)$ and $(B,C)$ at a depth\footnote{Defined as the distance of a node
from the root; depth of the root is 0.} of $i$ ($i > 0$) from $\T$'s root.
If we ignore the \ct and \bt terms in
\eqref{eqn:tree-rate}
, then the expected generation latency $T(i)$ of  both 
$(A,B)$ and $(B,C)$ being at depth $i$ is given by:
$T(i) = \frac{T}{2}(\frac{2}{3} \bp)^i.$
 Also, note that only \textit{one} of the \epss 
$(A,B)$ or $(B,C)$ waits for
$T(i)$ time on an average. Thus, the expected waiting times for each of the
four\footnote{Note that qubit $B$ in $(A,B)$ is different from that in $(B,C)$.}
qubits is $T(i)/2$. 

Based on the above, we can now easily estimate the total  
waiting by a qubit $q$ (referred to as $q$'s \textit{age}) 
before it is destroyed in a swapping operation. 
Let $l(q)$ be the leaf, i.e.\ the link \eps, of $\T$ that contains the qubit $q$. 
Let $\T(q)$ be the maximal subtree in $\T$ 
such that $l(q)$ is either its right-most 
or left-most leaf. Note that $\T(q)$ is well defined for a tree $\T$ and a qubit $q$. 
Let $d(q)$ be the depth of the root of $\T(q)$ in $\T$, 
and let $d'(q)$ be the depth of $l(q)$ in the subtree $\T(q)$. See Fig. \ref{fig:age}.
The expected age $A(q)$ of $q$ can  be estimated as follows. 
Note that \textit{age} of $q$ is the total waiting by $q$ 
at each of $l(q)$'s ancestors in $\T(q)$; also note that at $\T(q)$'s root, 
the qubit $q$ is destroyed, and hence, $q$ does not age at any 
ancestor of $T(q)$'s root. 
It is easy to see that the expected age $A(q)$ is:
$$ A(q) = \left( \sum_{i = d(q)}^{d(q) + d'(q)} T(i)/2 \right) + (t_{ob} + t_p)$$ 
Above, the last term is the time spent by $q$ waiting for its link \eps 
to be established and is given by sum of optical-BSM ($t_{ob}$) and photon
transmission latency ($t_p$).
Note that the actual age of a qubit $q$ is some distribution with the 
above mean. We observe the following.

\begin{observation}
Given a swapping tree $\T$, let $\T_l$ and $\T_r$ be its left and right children.
If the atomic BSM probability \bp is $\leq$ 75\%, then 
the expected age of the right-most or left-most descendant of either 
$\T_l$ or $\T_r$ is greater than the expected age of any other qubit in the tree. 
\label{ob:dec}
\end{observation}

\para{DP Formulation with Decoherence/Age Constraint.}
If we assume the atomic BSM probability \bp $\leq$ 75\%, then we can design
a DP algorithm for the \spp problem with the decoherence constraint,
as follows.
Let $T[i, j, \hh, \hh_{ll}, \hh_{lr}, \hh_{rl}, \hh_{rr}, u_i, u_j]$ 
be the optimal latency from a swapping tree of height at most \hh, 
whose root's 
left (right) child's left-most and right-most descendants are at \underline{depths} of
(\textit{exactly}) $\hh_{ll}$ and $\hh_{lr}$ ($\hh_{rl}$ and $\hh_{rr}$), each of which 
is upper bounded by $\hh$.
Here, $u_i$
and $u_j$ parameters are as before.
Note that
$T[i, j, 1, 0, 0, 0, 0, u_i, u_j] = \frac{\gt\min(u_i, u_j)}{\gp^2\ep^2\php}.$
We have: 

\begin{align} \label{eqn:dp-dec} 
    T&[i, j, \hh, \hh_{ll}, \hh_{lr}, \hh_{rl}, \hh_{rr}, u_i, u_j]\\
    &= \min\big( \nonumber 
    \!\begin{aligned}[t]
        & T[i, j, \hh-1, \hh_{ll}, \hh_{lr}, \hh_{rl}, \hh_{rr}, u_i, u_j], \\
        &(\frac{3}{2} B + \ct + \bt)/\bp\big)
    \end{aligned}\\
w&here: \nonumber \\
     B &=  \min_{k,\ g_i's, u + u' = 100} \max \big( \nonumber 
    \!\begin{aligned}[t]
        &T[
        \!\begin{aligned}[t]
            &i, k, \hh-1, \hh_{ll}-1, g_1, g_2\\
            &\hh_{lr}-1, u_i, u], 
        \end{aligned}\\
        &T[
        \!\begin{aligned}[t]
            &k, j, \hh-1, \hh_{rl}-1, g_3, g_4 \\
            &\hh_{rr}-1, u', u_j]\big)
        \end{aligned}
    \end{aligned}
t\end{align}

The above formulation will give us the optimal latency 
swapping-tree for each combination of 
$(\hh_{ll}, \hh_{lr}, \hh_{rl}, \hh_{rr})$. 
We remove the trees that violate the decoherence constraint, 
and pick the minimum-latency tree from the
remaining. This gives us a swapping tree with optimal latency 
under the decoherence constraint.
The proof of optimality easily follows.

\para{Constraint on Number of Leaves.}
Limiting the number of leaves to \fidl can be easily done by adding another parameter
for number of leaves \tqbl{in the $T$} array/function above. 
This adds another factor of $O(n^2)$ to the time 
complexity, as we need to check for \textit{all} 
combination of number of leaves in the
two subtrees. To optimize, we can now replace the height parameter, but keeping the
height parameters aids in parallelism, as described below.

\para{Time Complexity; \dpo and \dpa Algorithms}
Note that, in 
\eqref{eqn:dp-dec}
above, we can
pre-compute $\min_{g_1, g_2}$ $T[\ldots, g_1, g_2, \ldots]$
and similarly $\min_{g_3, g_4} T[\ldots, g_3, g_4, \ldots]$ before
computing $B$. With this, the time complexity of the DP formulation becomes $O(n^9)$, 
which can be further reduced $O(n^5 (\log n)^4)$ if we assume height of a tree to be at 
most $(c \log n)$ for some constant $c$. 
For a real-time routing application, the above time complexity is still high---as the algorithm
can take a few minutes on a single core. 
However, as the algorithm lends to obvious parallelism, it can be executed in as little as 
$O((\log n)^2)$ time with sufficiently many cores, using the height parameter sequentially.
We can also reduce the sequential time complexity to $O(n^5)$, by 
approximating the maximum qubit's age in a tree to the generation latency of 
the tree, which is a at most $3/(2\bp)$ the actual value. Note that maximum 
age of a qubit is at least $2T\bp/3$ and at most $T$, 
where $T$ is the generation latency of the tree.
Finally, we can make the algorithm more efficient
by assuming the usage parameter values 
to be 50\%.\footnote{This also enforces $s$ and $d$ to use only 50\% capacity;
this can be resolved by doubling  $s$ and $d$ capacity a priori.}
We refer to the $O(n^5)$ algorithm with the above assumptions as \dpa, and the 
$O(n^5 (\log n)^4)$ algorithm based on
\eqref{eqn:dp-dec}
as \dpo. Both algorithms
use throttling after the DP formulation.

\section{\dpalt Heuristic for \spp}
\label{sec:efficient}

The DP-based algorithms presented in~\S\ref{sec:single-path} for the \spp problem
have high time complexity, and thus, may not be practical for real-time route finding in large networks. In this section, we develop an almost-linear time 
heuristic for the \spp problem, based on the classic Dijkstra shortest
path algorithm; the designed heuristic performs close to the DP-based algorithms
in our empirical studies.

\para{Basic Idea.}
The main reason for the high-complexity of our DP-based algorithms
in~\S\ref{sec:single-path} is that
the goal of the \spp problem is to select an optimal swapping \textit{tree} 
rather
than a path. One way to circumvent this challenge efficiently while still
selecting near-optimal swapping tree, is to restrict ourselves to only 
``balanced'' swapping trees. This restriction allows us to think in terms
of selection of paths---rather than trees---since each path has a unique\footnote{In fact, there can be multiple balanced trees over a path whose length is not a power of 2, but, since they differ minimally in our context,
we can pick a unique way of constructing a balanced tree over a path.}
balanced swapping tree.
We can then develop an appropriate path metric based on above, and design
a Dijkstra-like algorithm to select an $(s,d)$ path that has the optimal
metric value.
We note that Caleffi~\cite{caleffi} also proposed a path metric based on 
balanced swapping trees, but their metric, though accurate, only had a
recursive formulation without a closed-form
expression---and hence, was ultimately not useful in designing an efficient
algorithm. In contrast, we develop an approximate
metric with a closed-form expression, 
based on the "bottleneck" link, as follows.

\para{Path Metric $M$.}
Consider a path $P = (s, x_1, x_2, \ldots, x_n, d)$ from $s$ to $d$,
with links $(s,x1), (x1, x2), \ldots, (x_n, d)$ with given \eps latencies.
We define the path metric for path $P$, $M(P)$, as the 
\eps generation latency of a balanced swapping over $P$, 
which can be estimated as follows.
Let $L$ be the link in $P$ with maximum generation latency. 
If $L$'s depth (distance from the root) is the maximum in a throttled 
swapping tree, then we can easily determine the accurate generation 
latency of the tree. However, in general, $L$ may not have 
the maximum depth, in which case we can still estimate the tree's latency 
approximately, if the tree is balanced, as follows.
In balanced swapping trees, \textit{assuming} the maximum latency link $L$
to be at the maximum depth, gives us a constant-factor approximation 
of the tree's generation latency. Thus, let us assume $L$ to be at 
the maximum depth of a balanced tree over $P$; this maximum depth
is $d = \lceil(\log_2 |P|)\rceil$.
Let the generation latency of $L$ be $T_L$.
If we ignore the $\bt + \ct$ term in
Eqn.~\eqref{eqn:tree-rate}
, then, the generation latency of a throttled swapping tree can be 
easily estimated to $T (\frac{3}{2\bp})^d.$ 
The term $\bt + \ct$  can also be incorporated as follows. 
Let $T(i)$ denote the expected latency of the ancestor of $L$ 
at a distance $i$ from $L$. Then, we get the recursive equation:
$T(i) = (\frac{3}{2}T(i-1) + \bt + \ct) / \bp.$
Then, the path metric value $M(P)$ for path $P$ is given by 
$T(d)$, the generation latency of the tree's root at a distance
of $d$ from $L$, and is equal to: 
$$M(P) = T(d) =\pp^{d}T_L + [(\pp^{d}-1)/(\pp-1)](\bt + \ct)/\bp$$ where 
$\pp = 3/(2\bp)$ and $d = \lceil(\log_2 |P|)\rceil$.
The above is a $(1+3/(2\bp))$-factor approximation latency 
of a balanced and throttled swapping tree over $P$; this can be shown easily
using analysis from \S\ref{sec:dec}.

\para{Optimal Balanced-Tree Selection.}
The above path-metric $M()$ is a monotonically increasing function over 
paths, i.e., if a path $P_1$ is a sub-sequence of another path $P_2$, 
then $M(P_1) \leq M(P_2)$. Thus, we can tailor the classical Dijkstra's shortest path
algorithm to select a $(s,d)$ path with minimum $M(P)$ value, 
using the link's \eps generation
latencies as their weights. We refer to this algorithm
as \dpalt, and it can be implemented with a time complexity of 
$O(m + n \log n)$ using Fibonacci heaps, where $m$ is the number of edges 
and $n$ is the number nodes in the network. 

\para{Incorporating Fidelity Constraints.} 
Fidelity constraints in our path-metric based setting can be handled by essentially
computing the optimal path for each path-length (number of hops in the path) up to
$\fidl$, and then pick the best path among them that satisfies the fidelity constraints.
This obviously limits the number of leaves to \fidl and addresses the operations-based
fidelity degradation. The above also address the decoherence/age
constraint, since it is easy to see (from analysis in \S\ref{sec:dec}) 
that the age of a balanced swapping tree can be very closely 
approximated in terms of the latency and the number of leaves.
Now, to compute the optimal path for each path-length, we can use a simple dynamic
programming approach that run in $O(m\fidl)$ time where $m$ is the number of edges 
and $\fidl$ is the constraint on number of leaves.      
\section{\iter: Iterative \qnr Heuristic}
\label{sec:iterative}

The general \qnr problem can be formulated in terms of hypergraph flows and solved using LP (see Appendix~\ref{sec:multiple-path}).  Although polynomial-time and provably optimal, the LP-based approach has a very high time-complexity for
it to be practically useful. Here, we develop an efficient heuristic
for the \qnr problem by iteratively using an \spp algorithm.

\para{\iter Heuristic.}
To solve the \qnr problem efficiently, we apply the efficient \dpa algorithm
iteratively---finding an efficient  swapping tree in each iteration 
for one of the $(s_i, d_i)$  pairs. The proposed algorithm is similar to the
classical Ford-Fulkerson augmenting path algorithm for the max network flow
problem at a high level, with some low level and theoretical differences as
discussed below.
The iterative-\dpa algorithm for the \qnr problem consists of the following
steps:
\begin{enumerate}
    \item  
    Given a network, we compute 
    maximum \eps generation rates for each network link using
    Eqn.~\eqref{eqn:gen-link-rate}.
    Use these as weights
    on the link. 
    
    \item
    For each $(s_i, d_i)$ pair, use \dpa algorithm to find the optimal path $P_i$,
    under the capacity and fidelity constraints. 
    Consider a \textit{throttled} and balanced swapping tree $\T_i$ over $P_i$.  
    Let $\T^*$ be the swapping tree with highest generation rate; if this 
    rate is below a certain threshold, then quit.
    
    \item
    Construct a residual network graph by subtracting the resources used by $\T^*$,
    using Eqn.~\eqref{eqn:node-res}.
    
    \item
    Go to step (1).
\end{enumerate}
\eat{Before we present the expressions required in the above algorithm, we make a few 
remarks. First, above, instead of \dpa, we can also use the
optimal \dpo algorithm, but as \dpo has high time-complexity, it may defeat
the whole purpose of designing an efficient algorithm in lieu of the optimal
LP formulation. We evaluate an iterative-\dpo approach in \S\ref{sec:eval}.}
Before we present the expressions required above, 
we would like to point out key differences of our context with the 
classic network flow setting. Even though we are augmenting our solution
one \textit{path} at a time, the network resources are fundamentally being used
by swapping trees created over these paths. These path-flows don't
really have a direction of flow, but we can assign them a symbolic direction 
from source to the direction. Even with these symbolic directions, the flows 
in opposite directions over any edge $k$ do not ``cancel'' each other as in
the classical network flow. Moreover, flow conservation law 
doesn't hold in our context (e.g., even a path may not use same link rates 
on all links, due to them being at different depths of the tree), and thus, 
the max-flow min-cut theorem doesn't hold.
Thus, \iter may not give an optimal solution,
even for a single $(s,d)$ pair.  

\para{Link \eps Generation Rate/Latency.} 
Consider a pair of network node $i$ and $j$ with corresponding \textit{current (residual)} 
values of node latencies as $\gt(i)$ and $\gt(j)$. 
Assuming \gp values to be same for both nodes,  
the minimum \eps link rate for $(i,j)$ is then given by

\begin{equation}
    \min(1/\gt(i), 1/\gt(j))\gp^2\ep^2\php.
    \label{eqn:gen-link-rate}
\end{equation}

\para{Residual Node Capacities.}
Let $P$ be a path added by \iter, at some earlier stage, and let 
$\T$ be the corresponding throttled swapping tree over $P$.
As in \S\ref{sec:problem}, let $R(e, \T)$ be the \eps generation
rate being used by \T over a link $e \in P$, 
$R_e = \sum_\T  R(e,\T)$, and $E(i)$ be the edges incident on $i$.
Then, the residual node rates can be calculated similar to
Eqn.~\eqref{eqn:qnr-1}
as follows. Below, $\gt'(i)$
is the \textit{original} value. 
\begin{eqnarray}
1/\gt(i) = 1/\gt'(i) -  \sum_{e \in E(i)} R_e/(\gp^2\ep^2\php) \ \ \ \  \forall i \in V. \label{eqn:node-res}
\end{eqnarray}
The residual memory capacity is easy to compute---each path/tree uses 2 memory units for each
intermediate node, and 1 memory unit for the end nodes. 

\eat{
\begin{equation}
1/\gt(i) = 1/\gt'(i) - 	\left(\sum_{(i,k) \in \P} R_{(i,k)} +   \sum_{(i,k) \in \P} R_{(i,k)}\right)/(\gp^2\ep^2\php)
    \label{eqn:node-res}
\end{equation}
\begin{equation}
1/\et(i) = 1/\et'(i) -  \left(\sum_{(i,k) \in \P} R_{(i,k)} +   \sum_{(i,k) \in \P} R_{(i,k)}\right) /(\red{\gp}\ep^2\php)
    \label{eqn:link-res}
\end{equation}
}

\section{Evaluations}
\label{sec:eval}

\begin{figure}
    \centering
    \includegraphics[width=0.45\textwidth]{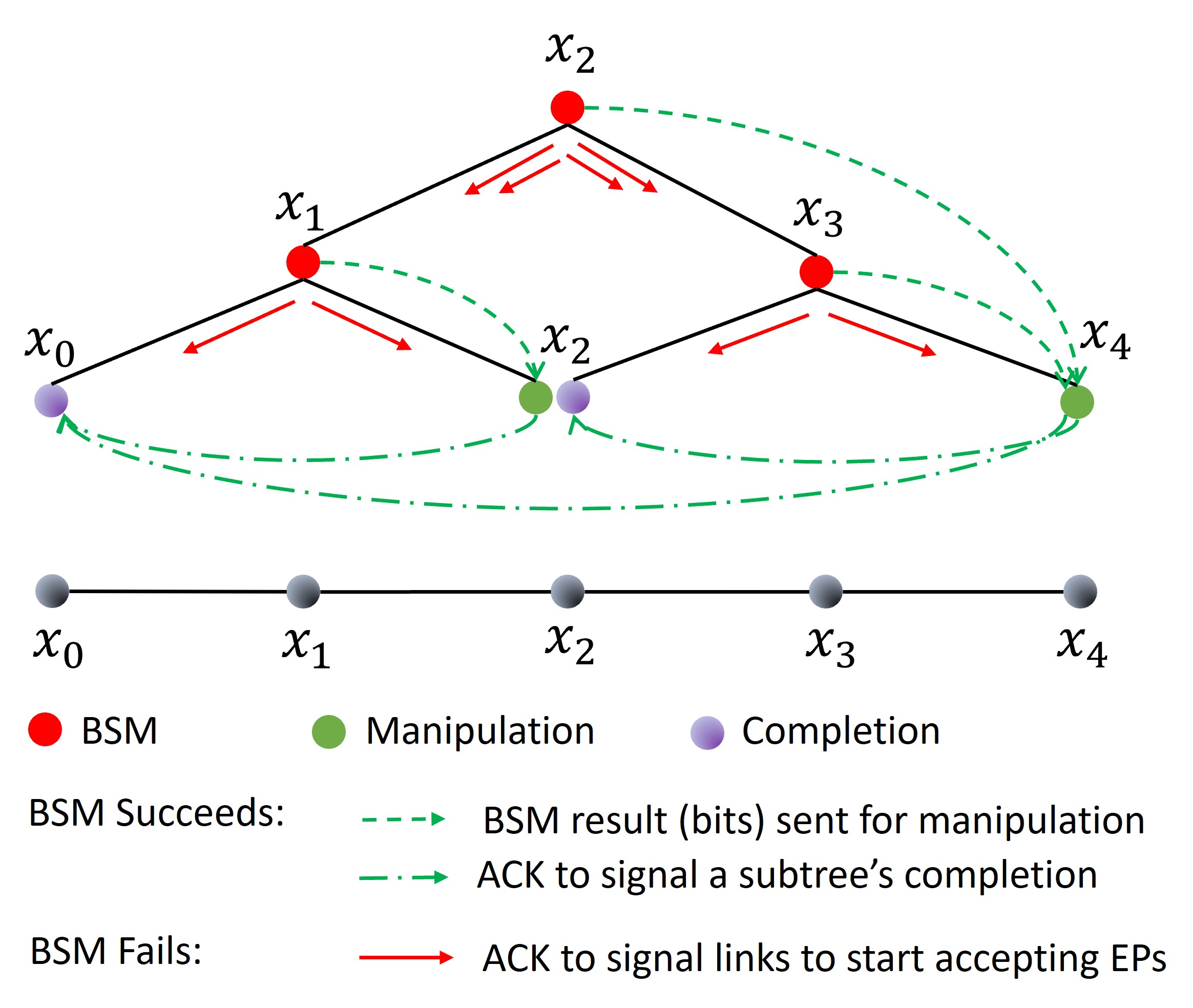}
    \vspace{0.1in}
  \caption{Swapping Tree Protocol Illustration. \red{The shown tree is not a swapping-tree, but rather a certain hierarchy of nodes to illustrate the BSM operation in the swapping-tree protocol.}  A link-layer protocol
  continuously generates \epss over links $(x_0, x_2)$ and $(x_2, x_4)$.
  On receiving \eps on links on either side, $x_1$ ($x_3$) attempts a BSM operation on the stored
  qubit atoms. If the BSM succeeds, $x_1$ ($x_3$) sends two classical bits (solid green arrows) to  $x_2$ ($x_4$) 
  for desired manipulation/correction after which $x_2$ ($x_4$) sends an ACK (dashed green arrows) to the other end-node $x_0$ ($x_2$) to complete the EP generation. If 
  BSM  at  $x_1$ and $x_3$ are both successful, then $x_2$ attempts the BSM as above. 
  If a BSM at say $x_1$ fails, that $x_1$ failure signals (red arrows) to all the descendant nodes of the subtree
  rooted at $x_1$ so that they can start accept new \epss from the link layer protocol. 
  Note that here node $x_2$ plays multiple roles and hence appears at multiple places in the figure.}
  \label{fig:protocol}
\end{figure}

\begin{figure*}[t]
    \centering
    \includegraphics[width=\textwidth]{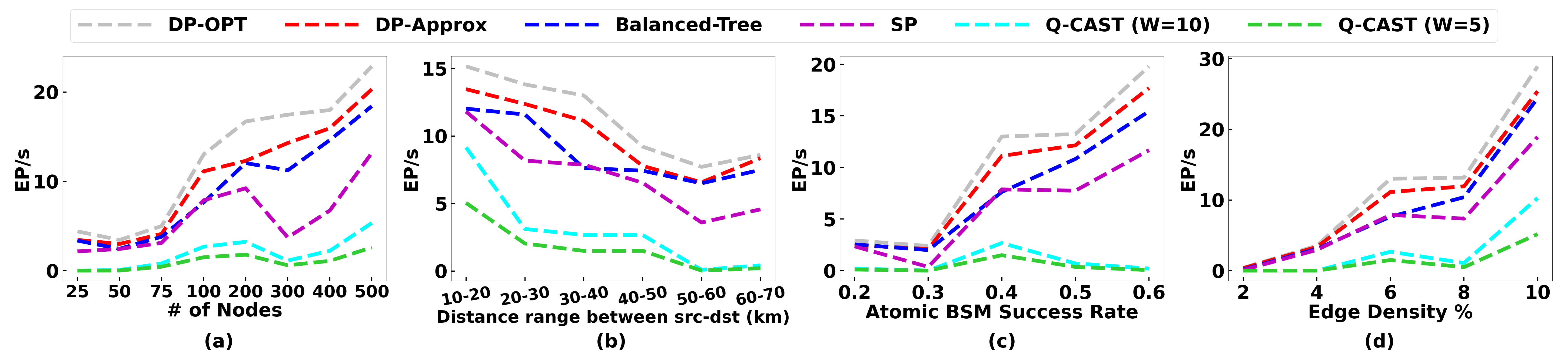}
    \caption{\spp Problem: \eps Generation Rates for varying parameters.}
    \label{fig:spp}
\end{figure*}

\begin{figure*}[t]
    \centering
    \includegraphics[width=\textwidth]{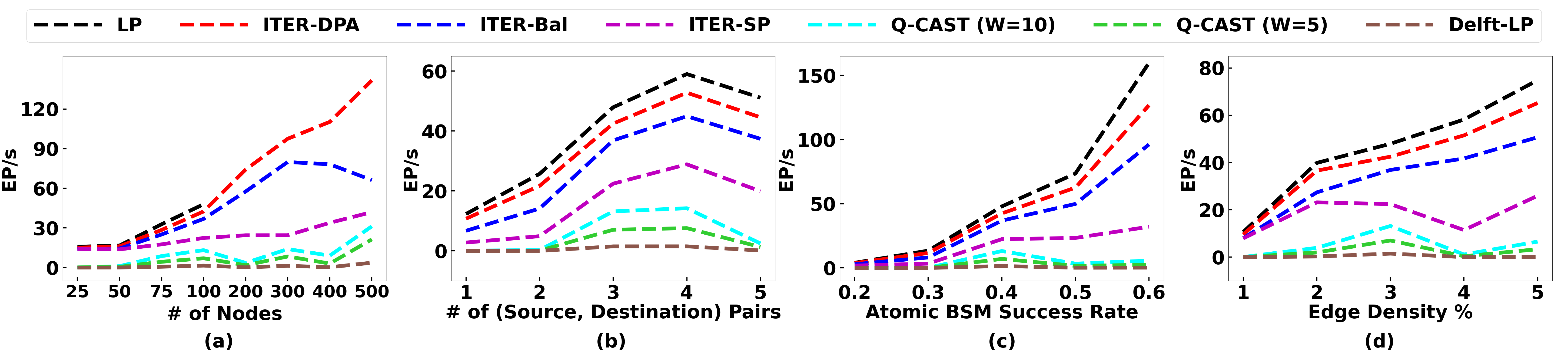}
    \caption{\qnr Problem: \eps Generation Rates for varying parameters.}
    \label{fig:qnr}
\end{figure*}

\begin{figure}[h]
    \centering
    \includegraphics[width=0.5\textwidth]{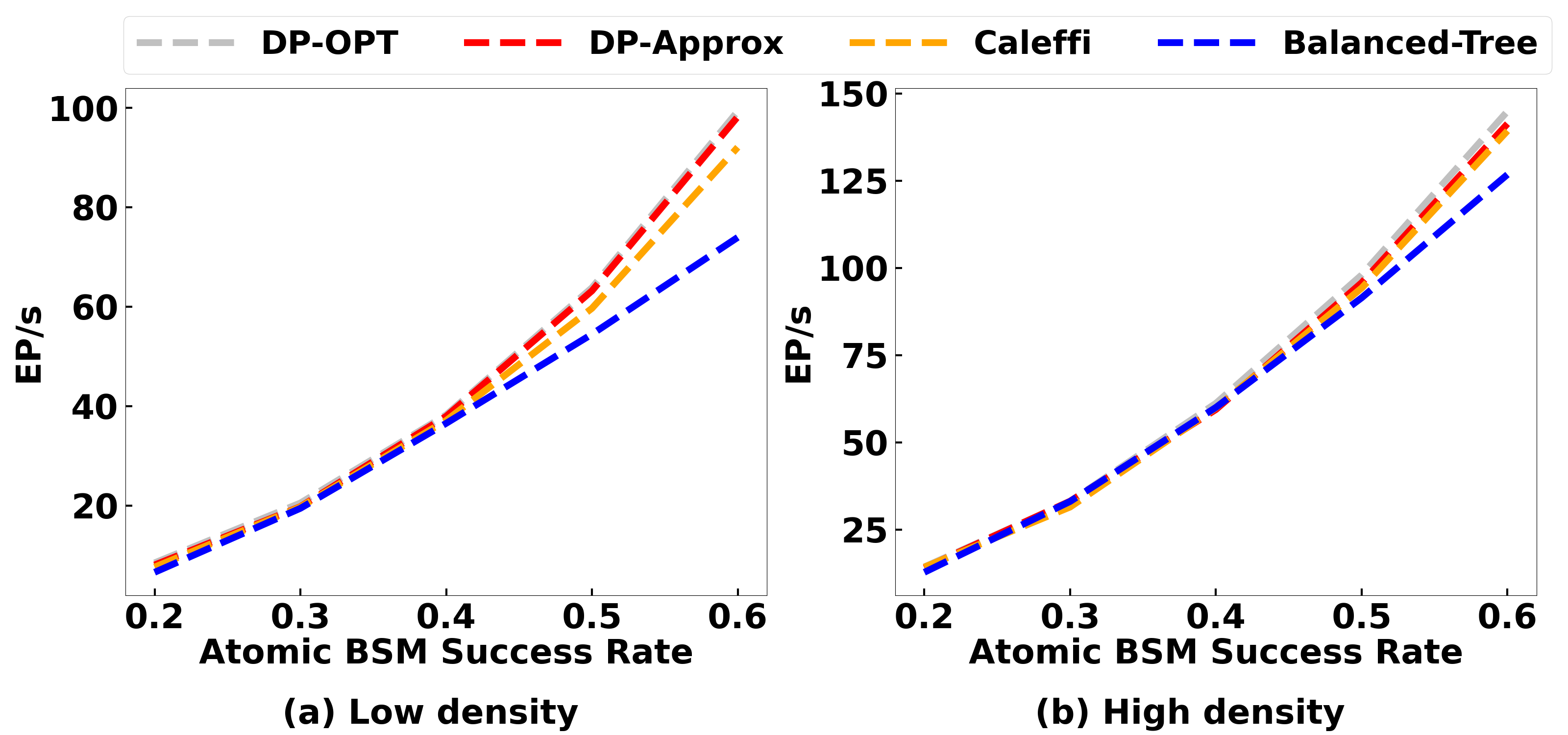}
    \caption{Compare the performance with Caleffi in a (a) low density network and a (b) high density network.}
    \label{fig:caleffi}
\end{figure}

The goal of our evaluations is to compare the \eps generation rates, evaluate
the fidelity of generated \epss, and validate our analytical models.
We implement the various schemes over a discrete event simulator 
for QNs called NetSquid~\cite{netsquid2020}. 
The NetSquid simulator accurately models various QN components/aspects, and
in particular, we are able to 
define various QN components and simulate swapping-trees protocols by
by implementing gate operations in entanglement swapping.

\para{Swapping Tree Protocol.}
Our algorithms compute swapping tree(s), and we need a way to implement
them on a network. 
We build our protocol on top of the link-layer of~\cite{sigcomm19},
which is delegated with the task of continuously generating \epss on a link at a desired rate (as per the swapping tree specifications).
Note that a link $(a,b)$ may be in multiple swapping trees, and hence, may need to handle multiple link-layer requests at the same time; we implement such link-layer requests by creating independent atom-photon generators at $a$ and $b$, with one pair of synchronized generators for each link-layer request. 
As the links generate continuous \epss at desired rates, we need a protocol to
swap the \epss. Omitting the tedious bookkeeping details, the key aspect of the protocol
is that swap operation is done only when both the appropriate \eps pairs have arrived.
We implement all the gate operations (including, atomic and optical BSMs) within 
NetSquid to keep track of the fidelity of the qubits. 
On BSM success, the swapping node transmits classical bits to the end node which manipulates its qubit, and send the final ack to the other end node. 
On BSM failure, a classical ack is send to all descendant link leaves, so that they can now start accepting new link \epss; note that in our protocol, a link $l$ does not 
accept any more \epss, while its ancestor is waiting for its sibling's \eps. See Fig.~\ref{fig:protocol}

\para{Simulation Setting.}
We use a similar setting as in the recent work~\cite{sigcomm20}.
By default, we use a network spread over an area of $100 km \times 100 km$.
We use the Waxman model~\cite{waxman}, used to create Internet topologies,
to randomly distribute the nodes and create links; we use the maximum link
distance to be 10km. We vary the number of nodes from \blue{25} to 
500, with 100
as the default value. We choose the two parameters in the Waxman model to
maintain the number of links to 3\% of the complete graph (to ensure an 
average degree of 3 to 15 nodes).
For the \spp problem, we pick $(s,d)$ pairs within a certain range of
distance, with the default being 30-40 kms; for the \qnr problem, we 
extend this range to 10-70 kms.


\softpara{Parameter Values.}
\blx{We use parameter values mostly similar to the ones used in~\cite{caleffi} corresponding to a single-atom based quantum memory platform, and vary some of them.}
In particular, we use atomic-BSM probability of success (\bp) 
to be 0.4 and latency (\bt) to be 10 $\mu$ secs; in some plots, we vary \bp from 
0.2 to 0.6. The optical-BSM probability of success (\php) is half of \bp. 
We use atom-photon generation times (\gt) and probability of success
(\gp) as 50 $\mu$sec and 0.33 respectively. Finally, we use photon 
transmission success probability 
as $e^{-d/(2L)}$~\cite{caleffi} where $L$ is the channel attenuation length
(chosen as 20km for an optical fiber) and $d$ is the distance between the nodes.
Each node's memory size is randomly chosen within a range of 15 to 20 units.
Fidelity is modeled in NetSquid using two parameter values, viz., depolarization
(for decoherence) and dephasing (for operations-driven) rates. 
We choose a decoherence time of two seconds based on achievable
values with single-atom memory platforms~\cite{loock20}; 
note that decoherence times of even several 
minutes~\cite{dec-13,dec-14} to hours~\cite{dec-15,dec-2021} has been 
demonstrated for other applicable memory platforms. 
Accordingly, we choose a depolarization rate of 0.01 such that the 
fidelity after a second is 90\%.
Similarly, we choose a dephasing rate of 1000 which corresponds to a link \eps 
fidelity of 99.5\%~\cite{delft-lp}.




\begin{figure*}
    \centering
    \includegraphics[width=\textwidth]{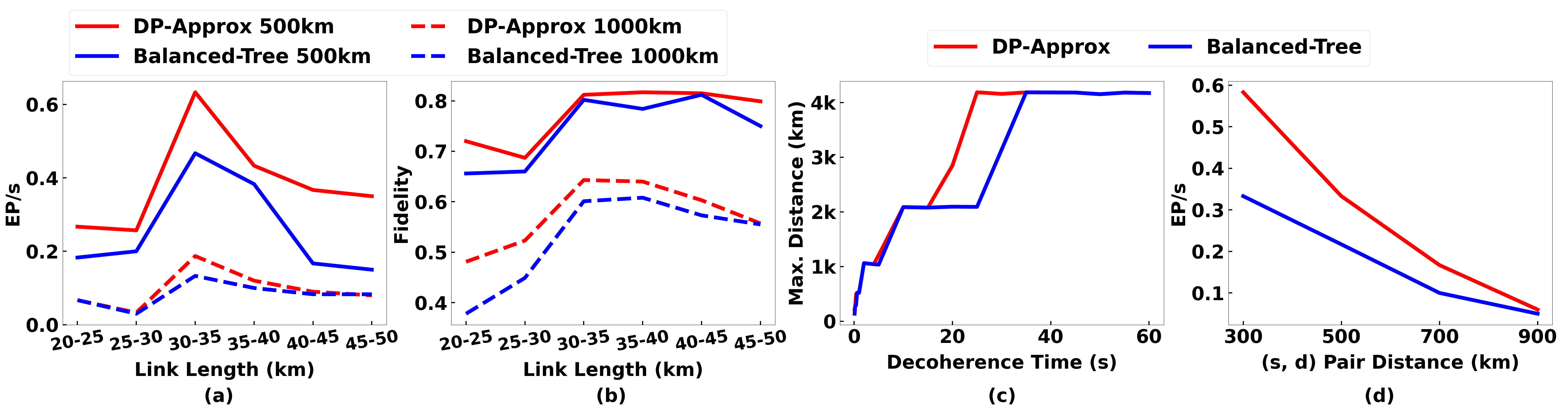}
    \caption{\eps generation over linear paths. 
    (a) \eps  Rates, and (b) Fidelity, over linear paths with varying link lengths. 
    (c) Maximum reachable distance with links of 30-35m lengths.
    (d) \eps generation rates over linear paths with 10-50 km links to demonstrate impact of varying link lengths.}
    \label{fig:fidelity}
\end{figure*}

\begin{figure*}
    \centering
    \includegraphics[width=\textwidth]{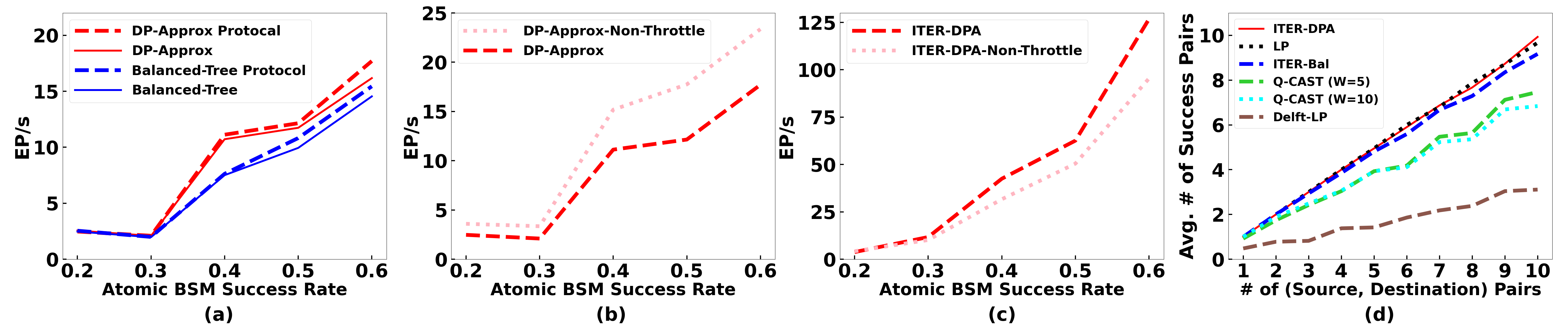}
    \caption{Comparing with Analytical Results. (a) Analytical vs.\ Simulation Results. (b) Throttled vs.\ Non-Throttled Trees (\spp). (c) Throttled vs.\ Non-Throttled Trees (\qnr). (d) Fairness Measure.}
    \label{fig:analytical}
\end{figure*}

\para{Algorithms and Performance Metrics.}
To compare our techniques with prior approaches, we implement most recently proposed
approaches, viz., (i) \blx{the \os-based \tqbl{linear programming (LP)} approach 
from~\cite{delft-lp} (called
\delftlp here), (ii) \qcast approach from~\cite{sigcomm20} which is \os-based but
uses multiple links and requires memories.} 
The \wt-based algorithm
by Caleffi~\cite{caleffi} uses an
exponential-time approach, and is thus compared only for small networks.
The~\cite{guha} and~\cite{greedy2019distributed} approaches are not compared
as they were found to be inferior to \qcast.

For all algorithm except for \qcast, we use only one link between adjacent nodes,
since only \qcast takes advantage of multiple links in a creative way. 
In particular, for \qcast, we use $W = 1, 5$, or 10 sub-links (\cite{sigcomm20} calls them
channels) on each link, with the node and link "capacity" divided equally among the 
them.
\eat{(e.g., for $W=5$ sub-links, each node's atom-photon generation capacity is divided among
$2W = 10$ sub-links to handle sub-links on either side).}
We note that in \qcast each node requires $2W$ memories (2 for each sub-link)
with sufficient coherence time to allow for the entire swapping operation over the
path to be completed.
The \delftlp approach explicitly assumes the generation of link \epss is deterministic,
i.e., the value $\gp^2\ep^2\php$ is 1, and does not model node generation rates. 
We address these differences by 
extending their LP formulation: (i) We add a constraint on node generation
rates, and (ii) add a $\gp^2\ep(i,j)^2\php$ factor to each link $(i,j)$ in any path
extracted from their LP solution.

Among our schemes, we use \dpo, \dpa and \dpalt (see \S\ref{sec:dec}) for the \spp problem, and \LP (Appendix~\ref{sec:multiple-path}) and \iter schemes for the \qnr problem. For \iter, we use three schemes:
\iterdpa, \iterdpalt and \iternaive, which iterate over \dpa, \dpalt and \naive respectively. To be comprehensive, 
we also implement a simple \naive algorithm which picks 
a balanced swapping tree over the shortest path (minimum number of links).
We compare the schemes largely in terms of \eps generation rates; we also
compare the execution times and \epss fidelity.

\para{Comparison with~\cite{caleffi} for \spp Problem.}
Note that~\cite{caleffi} gives only an \spp algorithm referred to as
\clf; it takes
exponential-time making it infeasible to run for
network sizes much larger than 15-20. In particular, for network sizes
17-20, it takes several hours, and our preliminary analysis suggests that
it will take of the order of $10^{40}$ \textit{years} on our 100-node
network. See Appendix~\ref{app:time}.
Thus, we use a small network of 15 nodes over a 25km $\times$ 25km area;
we consider average node degrees of 3 or 6. See Fig.\ref{fig:caleffi}.
We see that \dpo outperforms \clf by 10\% on a average for the sparser
graph and minimally for the denser graph. However, for some instances,
\dpo outperformed \clf by as much as 300\% (see Appendix~\ref{app:clf-perf}). We see that \dpa performs similar to \dpo,
while \dpalt is outperformed slightly by \clf; however, for this small
network, since the \dpo and \dpa algorithms only take 10-100s of
msecs (Appendix~\ref{app:time}), \dpalt need not be used in practice.  

\eat{
notably better than~
We see that each node has a degree of 3 on average, the performance of \dpa is almost 10\% better than Caleffi on average. We pick $(s,d)$, randomly. In cases, where even the best $(s,d)$ pairs' path are at least four hops and the difference between balanced vs. non-balanced comes to the picture, the performance difference between our \dpo (or even \dpa) than ~\cite{caleffi} is more substantial. See Fig.[??]. This difference can be as high as 3 times as Fig.~\ref{fig:non-balance} shows.
In Fig.~\ref{fig:caleffi}(b), a high density case, where each node has 6 neighbors, \dpa is still better than Caleffi, but the margin is smaller. 
The performance of \dpalt is inferior to Caleffi. 
}

\sloppypar
\para{\spp Problem (Single Tree) Results.}
We start with comparing various schemes for the \spp problem, in terms of \eps generation
rate. We compare \dpa, \dpo, \dpalt, \naive, and \qcast; 
note that the LP schemes can't be used to select
a \textit{single} tree, as they turn into ILPs. 
See Fig.~\ref{fig:spp}, where we plot the \eps generation rate for various schemes for
varying number of nodes, $(s, d)$ distance, \bp, and network
link density. We observe that \dpa and \dpo perform very closely, with the \dpalt heuristic
performing close to them; all these three schemes outperform the \qcast schemes (for $W=5, 10$ sub-links) by an 
order of magnitude. We don't plot \qcast for $W=1$ sub-links, 
as it performs much worse (less than $10^{-3}$ \eps/sec).
We note that \qcast's \eps rates here are much lower than the ones published in~\cite{sigcomm20}, because~\cite{sigcomm20} uses link \eps success 
probability of 0.1 or more, while in our
more realistic model, \blue{the link \eps success probability is 
$\gp^2\ep^2\php = 0.012$ for the default \bp value.}
We reiterate that our schemes require only 2 memory 
units per node, while the \qcast schemes requires $2W$ units. 
\eat{In addition, \qcast and \delftlp require 
tight synchronization of all links across a path, while
our schemes are asynchronous.}
The main reason for poor performance of \qcast (in spite of higher memory and link synchronization) 
is that, in the \os model, the \eps generation
over a path is a very low probability event---essentially $p^l$ where $p$ is the link-\eps success probability and $l$ is the path length, for the case of $W=1$
(the analysis for higher $W$'s is involved~\cite{sigcomm20}).
\blx{Finally, our proposed techniques also outperform the \naive algorithm, especially when the number of possible paths (trees) between $(s,d)$ pair increases.}
\blue{In addition, 
we see that performance increases with increase in \bp, number of nodes, or network link density, as expected due to availability of better trees/paths; it also increases with decrease in $(s,d)$ distance as fewer hops are needed.}

\para{\qnr Problem Results.}
We now present performance comparison of various schemes for the \qnr problem. Here, we compare
the following schemes: \iterdpa, \iterdpalt, \iternaive, \delftlp, and \qcast \bleu{with the optimal \LP as the benchmark for comparison (\LP wasn't feasible to run
for more than 100 nodes).}
See Fig.~\ref{fig:qnr}. Our observations are similar to that for the \spp problem results. 
We see that in all plots, \LP being optimal performs the best, but is closely matched by
\iterdpa and the efficient heuristic \iterdpalt.
\blx{We observe that  the performance gap between our proposed techniques and \iternaive is higher than in the \spp case, as \naive picks paths based on just number of 
links.}
Our schemes outperform both \delftlp and \qcast
by an order of magnitude, for the same reason as mentioned above. 

\para{Fidelity and Long-Distance Entanglements.}
We now investigate the fidelity of the \epss generated. 
First, we note that the \qcast and \delftlp schemes will incur near-zero decoherence 
as they involve only transient storage. 
Decoherence for other schemes is also negligible as the \eps generation 
latencies (10s of msecs) is much less than the coherence time. 
The operations-driven fidelity loss is expected to be similar for all schemes, as they
all roughly use the same order of links. 
Overall, we observed fidelities of 94-97\% across all schemes (not shown), 
with our schemes also
performing better sometimes due to smaller number of leaves.

\softpara{Long Path Graphs.}
To test the limits of the schemes in terms of decoherence and fidelity, 
we consider a long path network and estimate fidelity of \epss generated by
schemes for increasing distances and link-lengths (link success probability 
decreases with increasing link length).
Fig.~\ref{fig:fidelity}(a)-(b) shows \eps generation rates and fidelity for path lengths
of 500km and 1000km for varying link lengths, for the single-tree schemes \dpa and \dpalt. \qcast and \delftlp are not shown as their \eps rate is near-zero ($\leq 10^{-20}$) 
at these distances. 
We observe that our schemes yield \epss 
with qubit fidelities of 65-82\% and 40-64\% for 500km and 1000km paths respectively, 
with \eps rates of 0.05 to 0.65/sec. These are viable results---since 
\blue{qubit copies with fidelities higher than 50\% can be purified 
to smaller copies with 
arbitrarily higher fidelities~\cite{bennett-95,bennett-96}.}

\blx{
Now, in Fig.~\ref{fig:fidelity}(c), we demonstrate the effect of decoherence time of quantum memories used in nodes.
Here, we use 30-35 km links. We see that even with decoherence time of as low as 100 ms, \dpa is able to create \epss for up to 200 kms while \dpalt can only create \eps for paths up to~120 kms; they perform similarly for larger
decoherence times.
As all the links are almost of the same length, the optimal swapping will be largely-balanced trees wherein the \eps generation rate depends only on the 
tree height. Due to this reason, the maximum achievable path-length graph is 
close to a step function.
We add that our schemes produce 0.008 \epss/s for distance of more than 4000 kms.}

\blx{
Finally, in Fig.~\ref{fig:fidelity}(d), we demonstrate the higher performance of non-balanced trees when the links
on a path may have much different lengths. In particular, we pick link lengths randomly in the range of 10 to 50 kms. With this setting, we see that \dpa performs much better than \dpalt, and in some cases, up to 100\% better. 
Note that, \dpalt and \clf have similar performance over linear graphs, as there is no path selection scheme needed.}

\eat{
We observe that .. 
Fig.~\ref{fig:fidelity}(a) shows fidelity and \eps generation rates for varying
$(s,d)$ path lengths with link length in the range of 30-35 km, 
and Fig.~\ref{fig:fidelity}(d) shows fidelity and \eps generation rates for path length
of 500km with varying link length.}

\para{Validating the Analysis; Fairness.}
Fig.~\ref{fig:analytical}(a) compares the \eps generation rates as measured by the analytical formulae and the actual simulations for the \spp algorithms \dpa and \dpalt. We observe that they
match closely, validating our assumption of 3/2 factor in 
Eqn.~\eqref{eqn:tree-rate}
and of exponential distributions at higher levels of the tree, and of the path metric $M()$ for \dpalt.
Fig.~\ref{fig:analytical}(b)-(c) plots the \eps generation rates for throttled and non-throttled trees. We see that the throttled tree 
underperforms the non-throttled tree by only a small margin for the single-tree case; however,
for the multi-tree \iterdpalt algorithm,
the throttled trees perform better as they are able to use the resources efficiently. 
Fig.~\ref{fig:analytical}(d) plots the average number of $(s,d)$ pairs that get at least one
tree/path for varying number of requests; we see that our schemes exhibit 
90-99\% fairness.

\cbl
\para{Execution Times.}
We ran our simulations on an Intel i7-8700 CPU machine, and observed that  the \os algorithms as well our \dpalt and \iterdpalt
heuristics run in fraction of a second even for a 500-node network; thus,
they can be used in real-time. Note that since our problems depend on real-time network state (residual capacities), 
the algorithms must run very fast.
The other algorithms (viz., \dpo, \dpa, and \iterdpa) can take minutes to hours on large networks, and hence, may be impractical on large network 
without significant optimization and/or parallelization. See Appendix~\ref{app:time-plot} for the plot.
\cb

\eat{
\dpalt has the lowest runtime, with merely 0.0011 seconds in a 25 node network and 0.31 seconds in a 500 node network. 
\dpa takes from seconds to minutes and \dpo takes up to hours.
Fig.~\ref{fig:runtime}(b) shows the runtime of four algorithms for the \qnr problem.
\iterdpalt is the fastest and \iterdpa is the slowest.
In both \spp and \qnr problems, \qcast is only a little slower than our path heuristic.
\delftlp is slower than \qcast, and note that we used a package called OR-Tools by Google, which is written in C++ and is more efficient than our Python written code.
In summary, our path heuristic are the fastest and all the \os algorithms come next. 
And our DP based algorithms are the slowest. 
}

\section{Conclusions}
\label{sec:conc}

We have designed techniques for efficient generation of \eps to facilitate quantum
network communication, by selecting efficient swapping trees in a \wt protocol.
By extensive simulations, we demonstrated the effectiveness of our techniques,
and their viability in generating high-fidelity \eps over long distances (500-1000km).
Our future work is focused on exploring more sophisticated generation structures, 
e.g., aggregated trees, taking advantage of pipelining across rounds, incorporating
purification techniques, \tqbl{and to extend our techniques to multi-mode memories~\cite{simon07, collins07}.}

\section*{Acknowledgment}
This work was supported by \red{NSF awards FET-2106447 and CNS-2128187, and a Cisco industry grant.}
\appendices
\newcommand{\id}[1]{\ensuremath{\mathit{#1}}}
\newcommand{\tr}{\id{tr}}
\newcommand{\ttohp}{\pi}

\section{LP Formulation for the \qnr Problem}
\label{sec:multiple-path}

In this Appendix we provide an  optimal \LP-based solution to
the \qnr problem.  Although polynomial-time, this solution has high complexity, so its main use is as a benchmark in evaluating the more efficient (but possibly sub-optimal) algorithms for the problem.  

Our approach follows from the observation each
swapping tree in a QN can be viewed as a special kind of path (called
\emph{B-hyperpath}~\cite{Beckenbach2019}) over a hypergraph
constructed from the network graph.
%
We begin by describing the hypergraph construction for the
single-pair case and ignoring fidelity constraints. We then extend
traditional hypergraph-flow algorithm to incorporate losses
(e.g., due to BSM failures), stochasticity, and the interaction between 
memory constraints and stochasticity. Finally, we extend the formulation 
to multiple $(s,d)$ pairs and incorporate fidelity constraints.  

Optimal generation of long-distance entanglement was posed as an \LP
problem in~\cite{Daietal2020}, but differs from our \blue{more general formulation work} 
in three main
ways. First of all,~\cite{Daietal2020} assumes unbounded memory capacity at each
swapping node to queue up incoming \epss. In contrast, our model
has bounded memory capacity at each node, and consequently, 
our \LP formulation deals with \textit{expectations} over
rates/latencies rather than scalar rate values. 
Secondly, our formulation accounts for node capacity constraints in addition to link constraints. 
Thirdly, our formulation poses the problem in terms of hypergraph flows, which permits us to easily incorporate fidelity and decoherence constraints. 

\subsection{Hypergraph-Based Representation of Entanglement Generation}
We begin by recalling standard hypergraph notions~\cite{Beckenbach2019,GalloEtAl1993,ThakurTripathi2009}.

\begin{definition}{Hypergraph}
\label{defn:hypergraph}
\rm
  A directed hypergraph $H = (V(H),$ $E(H))$  has a set of vertices $V(H)$
  and a set of (directed) \emph{hyperarcs} $E(H)$, where each hyperarc $e$ is a
  pair $(t(e), h(e))$ 
  of non-empty disjoint subsets of $V(H)$.  
A \emph{weighted}
  hypergraph is additionally equipped with a weight function $\omega: E(H)
  \rightarrow R^+$.  
\end{definition}

Sets $t(e)$ and $h(e)$ are 
  called the \emph{tail} and \emph{head}, resp., of hyperarc $(t(e),
  h(e))$.
A hyperarc $e$ is a \emph{trivial} edge if both $t(e)$ and $h(e)$ are singleton; and \emph{non-trivial} otherwise. 
A hyperarc $e$ where $|h(e)| = 1$, i.e. whose head is singleton, is called a $B$-arc.
A hypergraph consisting only of $B$-arcs is called a $B$-hypergraph.

\begin{definition}{Connectivity and $B$-Hyperpaths}
\label{defn:hyperpath}
\rm
A vertex $t$ is \emph{$B$-connected to} vertex $s$ in hypergraph $H$ if
$s=t$ or there is a hyperarc $e \in E(H)$ such that $h(e) = \{t\}$ and
every $v \in t(e)$ is $B$-connected to $s$ in $H$.
A $B$-hyperpath from $s$ to $t$ is a minimal $B$-hypergraph $P$ 
such that $V(P) \subseteq V(H)$,  $E(P) \subseteq E(H)$, and $t$ is
$B$-connected to $s$ in $P$.
\end{definition}

\begin{figure}
    \centering
    \includegraphics[width=0.45\textwidth]{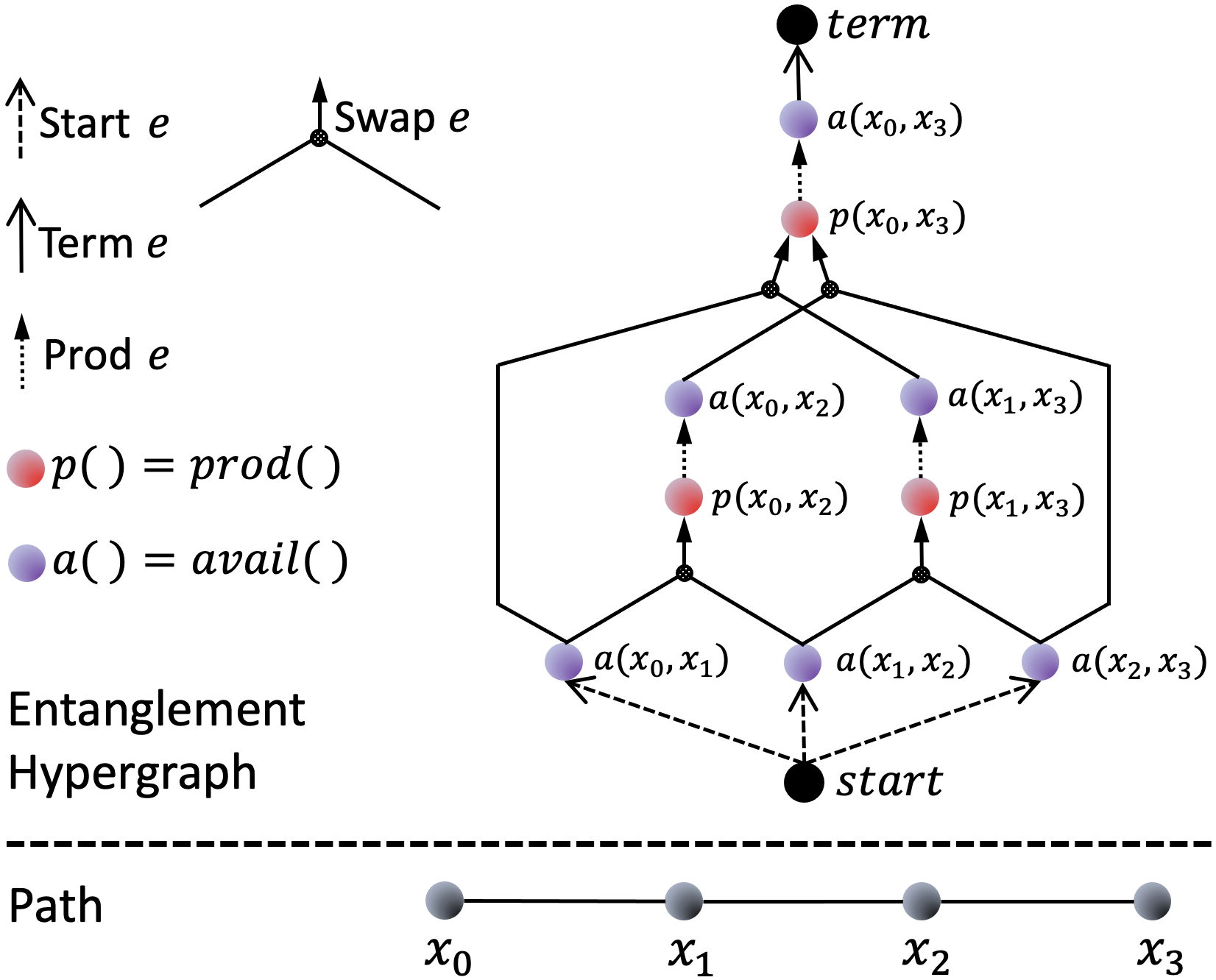}
    \vspace{0.1in}
    \caption{ST-hypergraph for a 4-node linear network. Not all $prod$ nodes are shown.}
    \vspace{-0.1in}
    \label{fig:LP}
\end{figure}

\para{ST-hypergraph.}
Given a QN and single $(s,d)$ pair, we first construct a hypergraph 
that represents the set of all possible swapping trees rooted
at $(s,d)$.
Given a QN represented as an undirected graph $G=(V,E)$
and a single $(s,d)$ pair, its \emph{ST-hypergraph} 
is a hypergraph $H$ constructed as follows  (see 
Fig. \ref{fig:LP}). All pairs of 
vertices below are unordered pairs.
\begin{itemize}
\item $V(H)$ consists of:
  \begin{enumerate}
  \item Two distinguished vertices $\id{start}$ and $\id{term}$
  \item $\id{prod}(u,v)$ and $\id{avail}(u,v)$ for all distinct $u, v \in
    V$
  \end{enumerate}
\item $\id{E(H)}$ consists of 5 types of hyperarcs:
  \begin{enumerate}
  \item\ [Start] $e = (\{\id{start}\}, \{\id{avail}(u,v)\}) \ \ \forall u,v \in V$.
  \item\ [Swap] $e = (\{\id{avail}(u,w),  \id{avail}(w,v)\}, \{\id{prod}(u,v)\})$, for all \emph{distinct} $u,v,w \in V$.
  \item\ [Prod] $e = (\{\id{prod}(u,v)\}, \{\id{avail}(u,v)\})\ \ \forall u,v \in V$.
  \item\ [Term]  $e = (\{\id{avail}(s,d)\}, \{\id{term}\})$.
  \end{enumerate}
 \end{itemize}
In an ST-hypergraph, vertices $\id{start}$ and $\id{term}$
represent source and sink nodes of a desired hypergragh-flow (see below).
Other vertices represent \epss over a pair of nodes in $G$.
Hyperarcs
represent how the tail \epss contribute to that at the head. 
For ease of accounting, we categorize generated \epss using different
types of vertices:
$\id{start}$ represents link-level \epss generated over links in $G$, 
$\id{prod}$ represent \epss produced by atomic entanglement-swapping, 
and $\id{avail}$ represent \epss generated from either of the above.
``Start'' and ``Prod'' arcs turn the $\id{start}$ and 
$\id{prod}$ \epss respectively into  $\id{avail}$ \epss and 
thus make them available for further swapping. 
``Swap'' arcs represent swapping over the triplets of nodes $(u,w,v)$.
Note that an ST-hypergraph is a $B$-hypergraph, as ``Swaps''
are the only non-trivial hyperarcs, and their head is singleton.

\softpara{Swapping Trees as $B$-Hyperpaths.} 
Given a \qnr problem with a single pair $(s,d)$, it is easy to see
that any swapping tree generating $(s,t)$ \epss can be represented 
by a unique $B$-hyperpath from \emph{start} to \emph{term} in the above 
ST-hypergraph. Thus, it easily follows that 
a \qnr problem of selection of (multiple)
swapping trees is equivalent to finding an optimal hypergraph flow 
from $\id{start}$ to $\id{term}$ in $H$. Note that $H$ has $O(|V|^2)$
vertices and $O(|V^3|)$ hyperarcs.

\subsection{Entanglement Flow as \LP}

We now develop an LP formulation to represent the \qnr problem over $(s,d)$ in $G$
as a hypergraph-flow problem in $H$.
In contrast to the classic hypergraph-flow formulation~\cite{Beckenbach2019}, 
we need to consider \emph{lossy} flow, with loss arising
from two sources: (i) \es operations have a given success probability, and (ii) waiting for both qubits to arrive before performing \es leads to losses since the arrival of \epss follow independent probability distributions.
For the latter, we make use of Observation~\ref{ob:expdist}.
The proposed LP formulation is as follows.
\begin{itemize}
\item \textbf{Variables}: $z_a$,  for each
  hyperarc $a$ in $H$, represents the \eps generation rate
  over each of the (one or two) node-pairs in $a$'s tail.
  This enforces the condition that \eps rates over the two 
  node-pairs in $\id{prod}$ hyperarc's tail are equal. Thus, the \LP 
  solution will result in \emph{throttled} swapping trees.
  
\item \textbf{Capacity Constraints}: $z_a \in R^+$ for all hyperarcs $a$ in $H$. 
We use 
Eqn.~\eqref{eqn:qnr-1}
to add the following constraints due to nodes in $G$. 
\begin{eqnarray*}
1/\gt &\geq& \sum_{x \in E(i)} z_{a(x)}/(\gp^2\ep^2\php) \quad  \forall i \in V.
\end{eqnarray*}
Above $a(x)$ is the hyperarc in $H$ of the form $(\id{start}, \id{avail}(x))$ 
where $x$ is an edge in $G$. 

\item \textbf{Flow Constraints} which vary with vertex types. Below, we use
notations $\id{out}(v)$ and $\id{in}(v)$ to represent outgoing and incoming
hyperarcs from $v$. Formally, $\id{out}(v)$ is $\{a \in E(H): v
\in t(a)\}$ and $\id{in}(v)$ is $\{a \in E(H): v \in h(a)\}$.

  \begin{itemize}
  \item For each vertex $v$ s.t. $v=\id{avail}(\cdot)$:
\[
  \sum_{a \in \id{in}(v)}  z_a = \sum_{a' \in \id{out}(v)} z_{a'}
\]
That is, there is no loss in making already generated entanglements
available for further swapping.

  \item For each vertex $v$ s.t. $v=\id{prod}(\cdot)$:
\[
  \sum_{a \in \id{in}(v)}  z_a\bp(2/3) = \sum_{a' \in \id{out}(v)} z_{a'}
\]
The $(2/3)\bp$ factor follows from Observation~\ref{ob:expdist}, and
accounts for loss due to swapping failures as well as due to waiting for
arrival of both \epss for swapping.
  \end{itemize}
\item \textbf{Objective}: Maximize $\sum_{a \in \id{in}(\id{term})} z_a$
\end{itemize}

\para{Multiple-Pairs Multi-Path:}
The above \LP formulation for the single-pair \qnr problem can be 
readily extended to the multiple-pairs case.
Let $\{(s_1, d_1), (s_2, d_2), \ldots, (s_n,d_n)\}$ be a set of
source-destination pairs. The \emph{only change} is
that the hypergraph $H$ now has $n$ arcs $(\{\id{avail}(s_i,d_i)\},
\{\id{term}\})$ for all $i$.  The other arcs model the generation of EPRs 
independent of the pairs, and thus are unchanged. 
It is interesting to note that the multi-pairs problem, 
typically formulated as multi-commodity flow in classical networks, 
is posed here as single-commodity flow over hypergraphs.

\subsection{Fidelity}

Constraints on loss of fidelity due to noisy BSM operations
and from decoherence due to the age of qubits can be added to the
\LP formulation, as follows.  
Recall that constraint on operation-based fidelity loss is modelled
by limiting the number leaves of the swapping tree, and in~\S\ref{sec:dec}, 
we formulated the decoherence constraint by limiting the
heights of the left-most and right-most descendants  of the root's
children. These
\emph{structural} constraints on swapping trees can be lifted to the
\LP formulation by adding the leaf count and heights as
parameters to $\id{prod}$ and $\id{avail}$ vertices; and (ii) swapping
the \epss generated from only the compatible vertices.  

In particular, we generalize the ST-hypergraph to a
\emph{fidelity-constrained} one called $H^{(F)}$, where the
$\id{prod}$ and $\id{avail}$ vertices are parameterized by $u,v \in
V$, and in addition by $(n,h)$ where $n$ is the number of leaves 
and $h=(h_{ll}, h_{lr}, h_{rl}, h_{rr})$ represents the depths of left-most 
and right-most descendants of the root's children, of the
trees rooted at $(u,v)$ with those parameter values.
In terms of edges, the most interesting difference $H^{(F)}$
and $H$ is in ``Swap'' edges.   In 
$H^{(F)}$, ``Swap'' edges are 
$( \{\id{avail}(u,w,n',h') $
$\id{avail}(w,v,n'',h'')\}, $
$\{\id{prod}(u,v,n,h)\})$ only if
$n = n'+n''$, and $h,h',h''$ are such that
$h_{ll} = h'_{ll}+1$, $h_{lr} = h'_{rr}+1$, $h_{rl}
= h''_{ll}+1$ and $h_{rr} = h''_{rr}+1$.  
The above constraints ensure that only compatible subtrees are
composed into bigger trees. The other changes are for bookkeeping: 
``Gen'' are from $\id{gen}(u,v)$ to $\id{avail}(u,v,1,(0,0,0,0))$; 
``Prod''  are from $\id{prod}(u,v,n,h)$ to
$\id{avail}(u,v,n,h)$; and finally ``Term''
are from $\id{avail}(s,t,n,h)$ to
$\id{term}$ for $n \leq \fidl$ and $h$ such that 
$f(h) \leq \fidd$; here, $f(h)$ gives the tree's age based on $h$ 
values (following \S\ref{sec:dec}) 
while using the link rates based on 50\% node-capacity 
usage.

\begin{table*}[t]
\centering
\caption{Execution times of \spp algorithm over small networks} 
\begin{tabular}{c rrrrrr} 
\hline\hline 
Algorithm&\multicolumn{6}{c}{Number of nodes} \\ [0.5ex]
& 10 & 13 & 15 & 16 & 18 & 20 \\
\hline 
\dpalt & 239$\mu$s & 360$\mu$s & 373$\mu$s& 492$\mu$s& 530$\mu$s& 552$\mu$s\\
\dpa & 4ms & 10ms & 14.7ms& 17.6ms& 28ms& 34ms\\
\dpo & 148ms & 363ms & 572ms & 706ms& 1s& 1.7s\\ 
Caleffi~\cite{caleffi} & 92ms & 4.6s & 14s& 26mins & 3.2hrs& 12.8hrs\\[1ex] 
\hline 
\end{tabular}
\label{tab:runtime}
\end{table*}

\newcommand*\pfofthm{PROOF OF THEOREM~\ref{thm:os-wt}}
\section{\protect\pfofthm}
\label{app:os-wt}

\para{Proof (sketch):} 
We provide a main intuition behind the claim in Theorem~\ref{thm:os-wt}.
The key claim is that at any instant the \os protocol generates an \eps,
the \wt protocol will also be able to generate an \eps.
Consider an instant $t$ in time when the \os protocol $X$ generates an 
\eps, as
a result of all the underlying processes succeeding at time $t$. 
Right before time $t$, consider the state of the \epss in the swapping-tree $T$ of the \wt 
protocol $Y$: Essentially, some of the nodes in $T$ have (generated) 
\epss that are waiting for their sibling \eps to be generated; note that these generated \epss have not aged yet, else they would have been already discarded by $Y$. Now, at time $t$, during $X$'s execution, all the underlying processes succeed instantly---it is easy to see that in the protocol $Y$ too, all the un-generated \eps would now be generated instantly\footnote{Here, we have implicitly assumed that if $n$ BSM operations succeed in $X$ protocol at some instant $t$, then at the same instant, $n$ BSM operations anywhere in $Y$ will also succeed.}---yielding a full \eps at the root (using qubits that have not aged beyond the threshold). Finally, since the number of operations in $T$ is the same as the number of BSM operations incurred by $X$ to generate an \eps, the fidelity degradation due to operations is the same in both the protocols.

\newcommand*\pfoflem{PROOF OF LEMMA~\ref{lem:subtrees}}
\section{\protect\pfoflem}
\label{app:subtrees}

\para{Proof.} We first prove the claim that given any swapping tree 
$\T_{xy}$ over a path  $P: x \leadsto w \leadsto y$, there exists
a swapping tree $\T_{xw}$ over a path $P':x \leadsto w$ such that 
$P'$ is a subset of $P$ and generation latency of $\T_{xw}$ is 
less than that of $\T_{xy}$. This claim can be easily proved by 
induction as follows. Consider two cases: (i) 
$w$ is the root of
$\T_{xy}$, in which case $T_{xw}$ is the left child of the root.
(ii) $i$ and $w$ have a common ancestor $a$ that is other than 
the root of $\T_{xy}$. In this case, $a$ = $w$, and the subtree 
rooted at $a=w$ is the required  $\T_{xw}$. (iii) The only common
ancestor of $i$ and $w$ is the root $a$ of $\T_{xw}$, which is not $w$.
In this case, we apply the inductive hypothesis on right subtree $\T_{ay}$
of
$\T_{xy}$, to extract a subtree $\T_{aw}$ which along with the left 
right subtree $\T_{ia}$ of $\T_{xy}$ --- gives the required subtree
$\T_{xw}$. This proves the above claim.

Now, to prove the lemma, let us consider the swapping trees
$\T_{ik}$ and $\T_{kj}$ given to us. By the above claim, there are 
swapping trees $\T_{iv}$ and $\T_{vj}$, which will satisfy the requirements
of the given lemma's claim.

\newcommand*\pfofthmt{PROOF OF THEOREM~\ref{thm:dp}}
\section{\protect\pfofthmt}
\label{app:dp}

\para{Proof.}
\tqbl{We show that $T[i, j, h, u_i, u_j]$ is indeed the optimal latency over the
nodes $(i,j)$ using a throttled swapping tree of height at most $h$ and 
with $u_j$ and $u_j$ as the usage percentages at nodes $i$ and $j$.
We use proof by induction over $h$.
The base case is obvious.
The inductive hypothesis is that the above statement is true for all heights $\leq (h-1)$.
Now, let $\T$ be an optimal-latency swapping tree of height at most $h$ between
a pair of nodes $(i,j)$, for some height greater than 1, and node usage percentages at $i$ and $j$ of $u_i$ and $u_j$ respectively. 
Let the expected latency of $\T$ be $L$. 
Let the two children subtrees of the root of $\T$ be $\T_1$ and $\T_2$, each of latency $L_c$; note that, as $\T$ is throttled, the expected latencies of  $\T_1$ and $\T_2$ are equal. Thus, we have $L_c = (\frac{3}{2}L + \ct + \bt)/\bp)$ by Eqn.~\eqref{eqn:tree-rate}.
Note that $\T_1$ and $\T_2$ are of heights at most 
$h-1$, and, without loss of generality, we can assume $\T_1$ and $\T_2$ to be disjoint (as per Lemma~\ref{lem:subtrees}).
Let $\T_1$ and $\T_2$ be between the pairs of nodes
$(i,k)$ and $(k,j)$ with end-nodes usage percentages of $(u_i, u_k)$ and $(u_k',u_j)$ respectively.
Now, optimal throttled trees over $(i,k)$ and $(k,j)$  must have a latency of
at most that of $\T_1$ and $\T_2$, i.e., $L_c$. 
Finally, by Eqn.~\ref{eqn:dp-usage} and the inductive hypothesis, we have that the 
$T[i, j, h, u_i, u_j]$ (and throttled) will be at most $L$. }

\newcommand*\exectime{EXECUTION TIMES OF \clf~\cite{caleffi} ALGORITHM}
\section{\protect\exectime}
\label{app:time}

Here, we give execution times of different algorithms especially \clf's for small networks of 10-20 nodes. See Table~\ref{tab:runtime}. We see that \dpalt and \dpa take fractions of a second, while \dpo takes upto 2 seconds. However, as expected \clf's execution time increases exponentially with increase in number of nodes -- with 20-node network takes 10+ hours. Below, we further estimate \clf's execution time for larger graphs.

\para{Rough Estimate of \clf's Execution Time for Large Graphs.}
Consider a $n$-node network with an average node-degree of $d$. Consider a node pair $(s,d)$. We try to estimate the number of paths from $s$ to $d$ -- the goal here is merely to show that the number is astronomical for $n$ = 100, and thus, our analysis is very approximate (more accurate analysis seems beyond the scope of this work). 
Let $P(l)$ be the number of simple paths from $s$ to a node $x$ in the graph of length at most $l$. For large graphs and large $l$, we can assume $P(l)$ to be roughly same for all $x$. We estimate that $P(l+1) = P(l) + P(l)*6*(1 - l/n)$. The first term is to count paths of length at most $l-1$; in the second term, the factor 6 comes from the fact the destination $x$ has 6 neighbors and the factor $(1-l/n)$ is the probability that a path counted in $P(l)$ doesn't contain $x$ (to constrain the paths to be simple, i.e,. without cycles). 
Using $P()$, the execution time of \clf can be roughly estimated to be at least $P(n-1)*500/(5*10^9)$ seconds where the factor 500 is a conservative estimate of the number of instructions used in computing the latency for a path and $5*10^{9}$ is the number of instructions a 5GHz machine can execute in a second.
The above yields executions times of a few seconds for $n=15$, 
about an hours for $n=20$,  
about 350 hours for $n=25$, 
and $10^{16}$ hours for $n=50$,
and $10^{44}$ hours for $n=100$. 
The above estimates for $n=15$ to 20 are within an order of 
magnitude of our 
actual execution times, and thus, validate our estimation approach.

\section{Comparison with \clf: More Details}
\label{app:clf-perf}

\begin{figure}[h]
    \centering
    \includegraphics[width=0.35\textwidth]{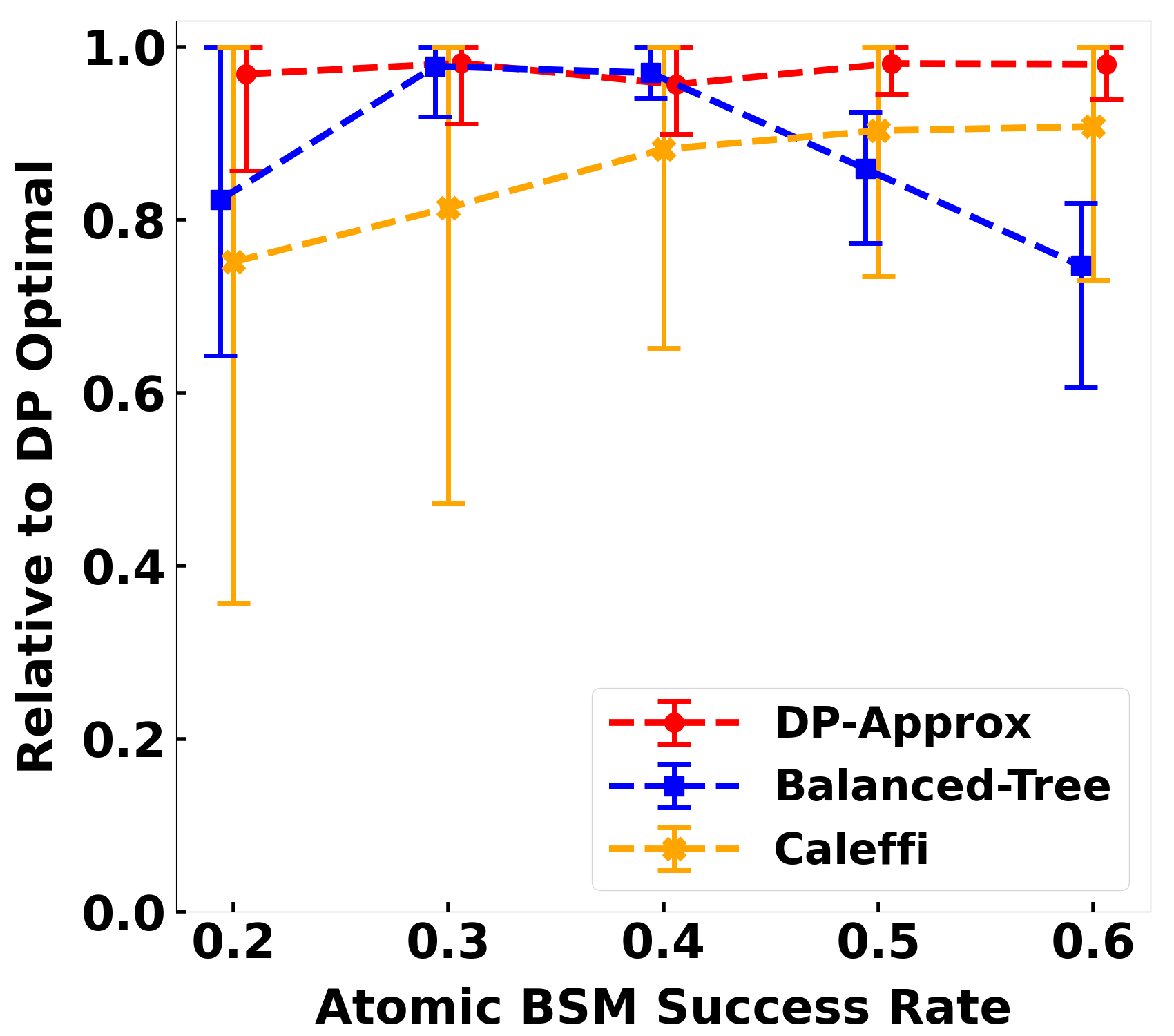}
    \vspace{0.05in}
    \caption{Compare the performance with Caleffi relative to \dpo (the closer to 1 the better).}
    \label{fig:caleffi-relative}
\end{figure}

Fig.~\ref{fig:caleffi} shows that \dpo outperforms \clf by a margin of around 10\% when averaging multiple experiments.
However, when we look at one experiment at a time and compute the \clf's performance relative to \dpo for each experiment, we see a larger difference between \dpo and \clf.
Fig.~\ref{fig:caleffi-relative} plots the error bar of the relative performance of three algorithms comparing to \dpo at each experiment.
The lower cap of \clf at 0.2 atomic BSM success rate is 0.35, which means that at an extreme sample, the \dpo is almost 300\% better than \clf.
In that extreme sample, the number of hops between the source and destination is large (thus the overall EP rate is small, which affects little when averaging with other experiments in Fig.~\ref{fig:caleffi}).
Moreover, we observe that the larger number of hops between the source and destination, the larger the gap of relative performance is between \dpo and \clf.
This observation aligns with what is shown in Fig.~\ref{fig:spp}(b): our \dpo has an larger advantage in ratio when the source and destination are far away.

\section{Execution Times Plot}
\label{app:time-plot}

\begin{figure}
    \centering
    \includegraphics[width=0.5\textwidth]{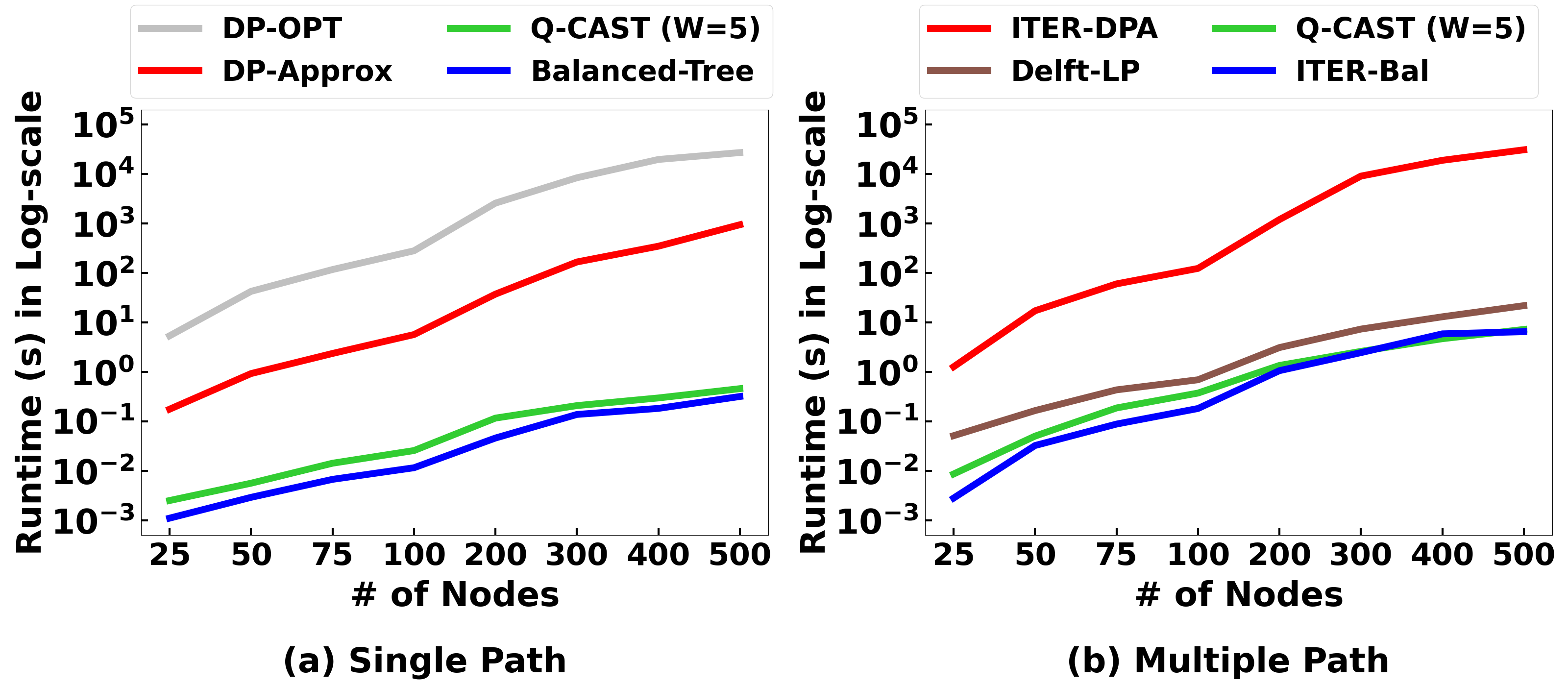}
    \vspace{-0.1in}
    \caption{The execution time comparison of various algorithms for \spp and \qnr algorithms.}
    \label{fig:runtime}
\end{figure}

We give here the plot for execution times of various schemes. 
See Fig.~\ref{fig:runtime}.

\bibliographystyle{IEEEtran}
\bibliography{routing, new}

\end{document}